
\input harvmac
\newcount\figno
\figno=0
\def\fig#1#2#3{
\par\begingroup\parindent=0pt\leftskip=1cm\rightskip=1cm\parindent=0pt
\global\advance\figno by 1
\midinsert
\epsfxsize=#3
\centerline{\epsfbox{#2}}
\vskip 12pt
{\bf Fig. \the\figno:} #1\par
\endinsert\endgroup\par
}
\def\figlabel#1{\xdef#1{\the\figno}}
\def\encadremath#1{\vbox{\hrule\hbox{\vrule\kern8pt\vbox{\kern8pt
\hbox{$\displaystyle #1$}\kern8pt}
\kern8pt\vrule}\hrule}}

\overfullrule=0pt

%
\def\underarrow#1{\vbox{\ialign{##\crcr$\hfil\displaystyle
 {#1}\hfil$\crcr\noalign{\kern1pt\nointerlineskip}$\longrightarrow$\crcr}}}
%

\def\bar{\overline}

\def\inbar{\vrule height1.5ex width.4pt depth0pt}
\def\IC{\relax\hbox{\kern.25em$\inbar\kern-.3em{\rm C}$}}
\def\IR{\relax\hbox{\kern.25em$\inbar\kern-.3em{\rm R}$}}
\def\IZ{\relax\ifmmode\hbox{Z\kern-.4em Z}\else{Z\kern-.4em Z}\fi}

\font\zfont = cmss10 

\def\bigone{\hbox{1\kern -.23em {\rm l}}}
\def\ZZ{\hbox{\zfont Z\kern-.4emZ}}


\def\drawbox#1#2{\hrule height#2pt
        \hbox{\vrule width#2pt height#1pt \kern#1pt
              \vrule width#2pt}
              \hrule height#2pt}

\def\Asym#1#2{\vcenter{\vbox{\drawbox{#1}{#2}
              \kern-#2pt       
              \drawbox{#1}{#2}}}}

\batchmode
  \font\bbbfont=msbm10
\errorstopmode
\newif\ifamsf\amsftrue
\ifx\bbbfont\nullfont
  \amsffalse
\fi
\ifamsf
\def\IR{\hbox{\bbbfont R}}
\def\IC{\hbox{\bbbfont C}}

\def\IZ{\hbox{\bbbfont Z}}


\midinsert
\endinsert


\nref\moha{R.N. Mohapatra, {\it Unification and Supersymmetry} second edition,
Springer-Verlag  (1992).}

\nref\alvarez{ E. Alvarez, ``Quantum Gravity: an Introduction to Some Recent Results'',
Rev. Mod. Phys. {\bf 61} (1989) 561.}

\nref\geom{ E. Witten, ``Physics and Geometry'', Proc. Int. Congress Math. 
(1986) pp. 267.}

\nref\gsw{ M. Green, J.H. Schwarz and E. Witten, {\it Superstring Theory},
Two volumes, Cambridge Univerity Press, Cambridge (1987).}

\nref\polbook{ J. Polchinski, {\it String Theory}, Two volumes,
Cambridge University Press, Cambridge (1998).}

\nref\lt{D. L\"ust and S. Theisen, {\it Lectures on String Theory},
Lectures Notes in Physics, Springer-Verlag, Berlin (1989).}

\nref\hatf{ B. Hatfield, {\it Quantum Field Theory of Point Particles and
Strings,} Addison-Wesley Publishing, Redhood City, (1992).}

\nref\oy{ H. Ooguri and Z. Yin, ``Lectures on Perturbative String
Theory'', hep-th/9612254.}

\nref\kirone{ E. Kiritsis, ``Introduction to Superstring Theory'', hep-th/9709062.}

\nref\shamit{ S. Kachru, ``Lectures on Warped Compactifications and Stringy Brane 
Constructions'', hep-th/0009247.}

\nref\ibanez{ F. Quevedo, ``Superstring Phenomenology: An
Overview'', Nucl. Phys. Proc. Suppl. {\bf 62} (1998) 134, hep-ph/9707434; 
L.E. Iba\~nez, ``The Second String (Phenomenology) Revolution'',
Class. Quantum Grav. {\bf 17} (2000) 1117; G. Aldazabal, L.E. Iba\~nez and F. Quevedo,
``On Realistic Brane Worlds From Type I Strings'', hep-th/0005033.}

\nref\fquevedo{ F. Quevedo, ``Lectures on Superstring Phenomenology'', in {\it Workshop
on Particles and Fields and Phenomenology of Fundamental Interactions} Eds. J.C. D'Olivo,
A. Fern\'andez and M.A. P\'erez, AIP (1996).}

\nref\neutrino{ E. Witten, ``Lepton Number and Neutrino Masses'', hep-ph/0006332.}

\nref\dine{ M. Dine, ``TASI Lectures on M Theory Phenomenology'', hep-th/0003175.}

\nref\pol{ J. Polchinski, ``TASI Lectures on D-Branes'', hep-th/9611050.}

\nref\cvj{ C.V. Johnson, ``D-brane Premier'', hep-th/0007170.}

\nref\banks{ T. Banks, ``Matrix Theory'', Nucl. Phys. Proc. Suppl. {\bf 67}
(1998) 180, hep-th/9710231; ``TASI Lectures on  Matrix Theory'', hep-th/9911068.}

\nref\taylor{ W. Taylor IV, ``The M(atrix) Model of M-Theory'', hep-th/0002016.}

\nref\gk{ A. Giveon and  D. Kutasov, ``Brane Dynamics and Gauge
Theory'', Rev. Mod. Phys. {\bf 71} (1999) 983, hep-th/9802067.}
  
\nref\malda{ S.S. Gubser, J.M. Maldacena, H. Ooguri and Y. Oz, ``Large $N$
Field Theories, String Theory and Gravity'', Phys. Rept. {\bf 323} (2000) 183,
hep-th/9905111.}

\nref\tsey{ A.A. Tseytlin, ``Born-Infeld Action, Supersymmetry and String Theory'',
hep-th/9908105.}

\nref\k{ E. Witten, ``D-branes and K-theory'',  JHEP {\bf 9812} (1998) 019,
hep-th/9810188.}

\nref\asen{ A. Sen, ``Non-BPS States and Branes in String Theory'', hep-th/9904207.}

\nref\kk{E. Witten, ``Overview of K-theory Applied to Strings'', hep-th/0007175.}

\nref\cube{ H. Garc\'{\i}a-Compe\'an and A.M. Uranga, ``Brane Box Realization of Chiral 
Gauge Theories in Two Dimensions'', Nucl. Phys. B {\bf 539} (1999) 329, hep-th/9806177.}

\nref\hw{ A. Hanany and E. Witten, ``Type IIB Superstrings, BPS Monopoles, and
Three-dimensional Gauge Dynamics'' Nucl. Phys. B {\bf 492} (1997) 152, hep-th/9611230.}

\nref\hz{ A. Hanany and A. Zaffaroni, ``On the Realization of Chiral Four-dimensional
Gauge Theories Using Branes'', JHEP {\bf 05} (1998) 001, hep-th/9801134.}

\nref\hu{ A. Hanany and  A. Uranga, ``Brane Boxes and Branes on Singularities'',
JHEP {\bf 9805} (1998) 013, hep-th/9805139.}

\nref\uranga{ A.M. Uranga, ``From Quiver Diagrams to Particle Physics'', hep-th/0007173.}

\nref\senlec{ A. Sen, ``An Introduction to Non-perturbative String Theory'',
hep-th/9802051.}

\nref\em{E. Witten, ``String Theory Dynamics in Various Dimensions'',
Nucl.Phys. B443 (1995) 85, hep-th/9503124.}

\nref\schwarz{ J.H. Schwarz, ``Lectures on Superstring and M Theory
Dualities'', Nucl. Phys. Proc. Suppl. {\bf 55B} (1997) 1, hep-th/9607201.}

\nref\town{ P.K. Townsend, ``Four Lectures on M-Theory, hep-th/9612121.}

\nref\vafa{ C. Vafa, Lectures on Strings and Dualities, hep-th/9702201.}

\nref\kirtwo{ E. Kiritsis, ``Introduction to Non-perturbative String Theory'',
hep-th/9708130.}

\nref\andreas{ B. Andreas, ``${\cal N}=1$ Heterotic/F-Theory Duality'', 
Fortsch. Phys. {\bf 47} (1999) 587, hep-th/9808159.}

\nref\ovrut{ B.A. Ovrut, ``${\cal N} =1$ Supersymmetric Vacua in Heterotic M-Theory'', 
hep-th/9905115.}

\nref\fmw{ S. Sethi, C. Vafa and E. Witten, ``Constraints on Low Dimensional 
String Compactifications'', Nucl. Phys. B {\bf 480} (1996) 213, hep-th/9606122; R.
Friedman, J. Morgan and 
E. Witten, ``Vector Bundles and F-theory'', Commun. Math. Phys. {\bf 187} (1997) 679,
hep-th/9701162.}



\Title{hep-th/0010046, CINVESTAV-FIS-99/33}
{\vbox{\centerline{Topics on Strings, Branes} 
\medskip
\centerline{and Calabi-Yau Compactifications}}\foot{ This survey
is dedicated to the memory of the father of H. G.-C., Sr. Mariano Garc\'{\i}a Salazar.}}
\smallskip
\centerline{Hugo
Garc\'{\i}a-Compe\'an\foot{E-mail address:
{\tt compean@fis.cinvestav.mx}} and Oscar
Loaiza-Brito\foot{E-mail address: {\tt oloaiza@fis.cinvestav.mx}}}
\smallskip
\centerline{\it Departamento de F\'{\i}sica}
\centerline{\it  Centro de Investigaci\'on y de Estudios Avanzados del IPN}
\centerline{\it Apdo. Postal 14-740, 07000, M\'exico D.F., M\'exico}

\bigskip  
\medskip
\vskip 1truecm
\noindent
Basics of some topics on perturbative and non-perturbative string theory are reviewed. 
After a mathematical survey of the Standard Model of particle physics and GUTs, the
bosonic string kinematics for the free case and with interaction is described. 
The effective action of the bosonic string and
the spectrum is also discussed. Five perturbative superstring theories and their spectra
is briefly outlined. Calabi-Yau three-fold compactifications of heterotic strings and their
relation to some four-dimensional physics are given. T-duality in closed and open strings 
are surveyed. D-brane definition is provided and some 
of their properties and applications to brane boxes configurations, in particular to 
the cube model are discussed. Finally, non-perturbative issues like S-duality, M-theory, 
F-theory and  basics of their non-perturbative Calabi-Yau compactifications are
considered.


\noindent

\Date{October, 2000}



\newsec{Introduction}

String theory is by now, beyond the Standard Model (SM) of particle physics, the best and
most sensible understanding of all of the basic components of matter and their interactions
in an unified scheme.  There are well known the `aesthetic' problems arising in the standard
model of particles, they include the hierarchy problem, the abundance of free parameters and
the apparent arbitrariness of the flavor and gauge groups. The SM is for this reason
commonly regarded as the low energy effective description of a more fundamental theory,
which solves these problems (for a nice review see \moha). It is also widely
recognized that Quantum Mechanics and General Relativity cannot be reconciliated in the
context of a perturbative quantum field theory of point particles.  Hence the
nonrenormalizability of the general relativity can be regarded (similarly to the standard
model case) as a genuine evidence that it is just an effective field theory and new physics
associated to some fast degrees of freedom should exist at higher energies (for a review,
see for instance \alvarez). String theory propose that these fast degrees of freedom
are precisely the strings at the perturbative level and at the non-perturbative level the
relevant degrees of freedom are higher-dimensional extended objects called D-branes (dual
degrees of freedom).

At the perturbative level String Theory has intriguing generic predictions such as:  $(i)$
Spacetime supersymmetry, $(ii)$ General Relativity and $(iii)$ Yang-Mills fields.  These
subjects interesting by themselves are deeply interconnected in a rich way in string theory. 

The study of theories involving D-branes is just in the starting stage and many surprises
surely are coming up. Thus we are still at an exploratory stage of the whole structure of
the string theory. Therefore the theory is far to be completed and we cannot give yet
concrete physical predictions to take contact with collider experiments and/or astrophysical
observations. However many aspects of theoretic character, necessary in order to make of
string theory a physical theory, are quickly in progress.  The purpose of these lectures are
to overview the basic ideas to understand these progresses. This paper is an extended
version of the lectures presented at the {\it Ninth Mexican School on Particles and Fields}
held at Metepec Puebla. M\'exico. We don't pretend to be exhaustive and we will limit
ourselves to describe the basics of string theory and some particular new developments like
Calabi-Yau compactifications of the M and F theories.  We apologize for omiting numerous
original references and we prefer to cite review articles and some few seminal papers.

We first survey very briefly some basic concepts of the gravitational field and gauge
theories pointing out the difficulties to put they together. After that we overview the
string and the superstring theories, including Calabi-Yau compactifications and the relation
of strings to physics in four dimensions from the perturbative point of view. T-duality,
$D$-branes and brane boxes configurations is also considered. After that, we devote some
time to describe the string dualities and the web of string theories connected by dualities.
M and F theories are also briefly described. Finally we overview an approach to
non-perturbative Calabi-Yau compactifications of M and F theories.

\vskip 2truecm
\newsec{Motivation for Using Strings}

First we overview the basic structure of General Relativity (GR) and Yang-Mills (YM) theory
in four dimensions. They are very different theories. GR for instance, is the dynamical
theory of the spacetime metric while quantum YM theories and in general, Quantum Field
Theory (QFT)  describes the dynamical building blocks of matter in a fixed spacetime
background. Here we survey basics aspects of GR and YM theory following closely Ref.
\geom.

The
pure gravitational field is described by a pseudo-Riemannian metric $g_{\mu \nu}$ 
with $\mu , \nu =0,1,2,3$ (on a
four-dimensional manifold $M$) satisfying the
vacuum Einstein equations, $R_{\mu \nu}=0$. Einstein equations can be derived from the
Einstein-Hilbert action

\eqn\uno{
S_{GR} = {1 \over 16 \pi G_N} \int_M d^4x \sqrt{-g} R,}
where $G_N$ is the Newton's constant. This constant together with $\hbar$ and $c$, deremines the
Planck scale where gravitational effects in the quantum theory are relevant. The mass scale
termed Planck mass is $M_{Pl} = \sqrt{\hbar c \over G_N}= 1.2 \times 10^{-5} grams$ or equivalently
the Planck length $L_{Pl} = {\hbar \over M_{Pl} \ c} \approx 10^{-33} centimeters.$ 

On the other hand the SM of particle physics is described by
the gauge field theory which is a quantum field theory provided with the gauge symmetry
structure.
If one wants to formulate the gauge theory on a pseudo-Riemmanian manifold spite of the
metric $g_{\mu \nu}$ we require from an additional structure on the spacetime {\it i.e.} a connection
$A$ on a $G$-principal bundle on $M$: $G \to E \to M$, where $G$ is the SM gauge group,
$G = {\rm SU}_C(3) \times {\rm SU}_L(2) \times {\rm U}_Y(1)$. As usual, the gauge field
$A_{\mu}(x)$ given by
the connection one-form has associated the field strength $F_{\mu \nu} = \partial_\mu A^a_\nu -
\partial_\nu A^a_\mu + i f^a_{bc} A^a_{\mu} A^b_{\nu}$ with $f^a_{bc}$ being the structure 
constants of $G$. 
Given any representation ${\cal R}$ of $G$ one can construct the {\it associated vector bundle}
$V_{\cal R}$. 

The Yang-Mills action is given by

\eqn\dos{
S_{YM} = - {1 \over 4 g^2_{YM}} \int_M g^{\mu \mu '}g^{\nu \nu '} Tr_{\cal R} F_{\mu \nu}
F_{\mu ' \nu ',}}
where $Tr_{\cal R}$ denotes the trace in the adjoint representation of $G$.

Now we would like to introduce fermions. The chiral fermions are sections of the 
the chiral spin bundles $\widehat{S}_{\pm}$ over spacetime manifold with $Spin$ structure
$M$, {\it i.e.} 
$\widehat{S} \to M$, where $\widehat{S} = \widehat{S}_+ \oplus \widehat{S}_-$. The
fibers are the Clifford modules constructed with the Dirac matrices $\Gamma^{\mu}$. 
Dirac operator is $ \not \! \! D \equiv \Gamma^{\mu} D_{\mu}: \Gamma(\widehat{S}) \to
\Gamma(\widehat{S})$ with $D_{\mu}$ being the spacetime covariant derivative. 
In even dimensions Dirac operator decomposes as: $\not \! \! D =\not \! \! D_+ \oplus \not \!
\! D_-$ where 
$\not \! \! D_{\pm}:
\Gamma(\widehat{S}_{\pm})
\to \Gamma(\widehat{S}_{\mp})$ with

\eqn\tres{
\not \! \! D_- \psi_+ = 0, \ \ \ \ \ \ \ \ \  \ \not \! \! D_+ \psi_- = 0,}
and $\psi_{\pm} \in \Gamma(\widehat{S}_{\pm})$. 

The possibility to add a mass term in the above equation implies that that mass should be 
of order one in mass Planck unities $M_{Pl}$. But the mass of the low energies particle $m$ 
should be much lower than the Planck mass $M_{Pl}$, {\it i.e.} $m << M_{Pl}$. A very
nice solution can be 
given by introducing gauge fields in {\it complex} representations of the gauge group $G$.
In that case  the fermions should be sections of the original spin bundle
$\widehat{S}$ coupled to the associated vector bundle
$V_{\cal R}$.
If ${\cal R} \not\cong \widetilde{\cal R}$ then the corresponding bundles are not isomorphic 
$V_{\cal R}  \not\cong V_{\widetilde{\cal R}}$. So we have four possibilities

$$ 
W_+ = \widehat{S}_+ \otimes V_{\cal R}, \ \ \ \ \  W_- = \widehat{S}_- \otimes V_{\cal R},
$$

$$ 
\widetilde{W}_+ = \widehat{S}_+ \otimes V_{\widetilde{\cal R}}, \ \ \ \ \ \ 
\widetilde{W}_- = \widehat{S}_- \otimes V_{\widetilde{\cal R}}.
$$
$CPT$ theorem implies that the fermions with different chirality are given by

$$ \psi_+ \in \Gamma(\widehat{S}_+ \otimes V_{\cal R}), \ \ \ \ \ \ \ 
   \widetilde{\psi}_- \in \Gamma (\widehat{S}_- \otimes V_{\widetilde{\cal R}}).
$$
This explains why the mass term are not allowed in the right-hand of Eq. (2.3).

Now define ${\cal R} = \oplus_{i=1}^{15} {\cal R}_i$ and  $\widetilde{\cal R} =
\oplus_{i=1}^{15} \widetilde{\cal R}_i$ where ${\cal R}_i$ and  $\widetilde{\cal R}_i$
are irreducible complex representations of the gauge group of the SM. Define the formal 
difference 

\eqn\cuatro{
\Delta \equiv U \ominus \widetilde{U}}
between the general complex representation of a particle $U = U_0 \oplus {\cal R}$ and
its corresponding complex conjugated $\widetilde{U}= U_0 \oplus \widetilde{\cal R}$
with $U_0$ being a {\it real} irreducible representation (irrep). Thus in the computation of
$\Delta$
only the complex representations are relevant, {\it i.e.} $\Delta = {\cal R} \ominus 
\widetilde{\cal R}$.

On the other hand, spontaneously symmetry breaking is then the important  mechanism to give
mass to the
fermions and gauge particles in the SM. The possibility to get lower symmetries through
the breaking of the gauge group lead us to consider theories with
higher symmetries than the SM and recuperate it by symmetry breaking. 
These are the Grand Unified Theories (GUTs). The extension of the gauge 
group $G$ to another of higher dimensionality $\bar{G}$ was an exciting hope for understanding 
the `aesthetic' problems of the SM mentioned in the 
introduction.  One of the more successful GUT is the so called SU(5) GUT, where the  gauge
group $\bar{G}$ is SU(5) and it breaks to the SM group. Computation of $\Delta$ for this
model consist in taking the formal difference between all irreducible representations of 
SU(5) and their complex conjugated ones, this gives

\eqn\cinco{
\Delta = 3 \bigg( {\bf 5}^* \oplus {\bf 10} \ominus 
{\bf 5} \oplus {\bf 10}^* \bigg),}
where ${\bf 5}$ and ${\bf 5}^*$ are the fundamental and anti-fundamental representations of
SU(5),  and  ${\bf 10}$ and ${\bf 10}^*$
are the antisymmetric part of representation ${\bf 5} \otimes {\bf 5}$  of
SU(5) and its complex conjugated. The `3' in the front part stands for the mysterious number of
generations of quarks and leptons. We will come back later to comment about this mysterious 
number.

SU(5) is by itself a non-trivial maximal subgroup of SO(10). The GUT with gauge group SO(10)
is another candidate
for a unified model.
The decomposition of irreps of SO(10) in terms of irreps of SU(5) is as follows: the
fundamental representation of SO(10) ${\bf 10}$ decomposes under SU(5) irreps as ${\bf 10} =
{\bf 5} \oplus {\bf 5}^*$. SO(10) has
two complex conjugated spinor representations of 16 dimensions, they are: ${\bf 16}$ and
${\bf 16}^*$. They can be decomposed under SU(5) irreps  as, ${\bf 16} =
{\bf 1} \oplus {\bf 5}^* \oplus {\bf 10}$ and ${\bf 16}^* =
{\bf 1} \oplus {\bf 5} \oplus {\bf 10}^*$. Then computation of $\Delta$ yields

\eqn\seis{
\Delta=  3 \bigg( {\bf 16} \ominus {\bf 16}^*\bigg).}

Higher dimensionality group $E_6$ is the next candidate for a GUT. This group has 
complex representations which are: ${\bf 27}$ and ${\bf 27}^*$. Under SO(10) irreps, these
representations decompose into the spinor, vector and identity irreps: {\it i.e.} ${\bf 27} =
{\bf 16} \oplus {\bf 10} \oplus {\bf 1}$. Vector representation is real. Thus $\Delta$ is
computed easily to get

\eqn\siete{
\Delta = 3 \bigg( {\bf 27} \ominus {\bf 27}^*\bigg).}
Bigger exceptional groups like $E_8$ only has real representations and $\Delta = 0$.

The SM and GUTs are thus unable to answer the arbitrariness of the number of families of lepton and
quarks (basically the `3' arising in Eqs. (2.5), (2.6) and (2.7)) as well as the arbitrariness of
the gauge group. The hierarchy of lepton and quarks masses, the existence of the Higgs mechanism and
the abundance of free parameters
are `aesthetic problems' as they don't contradict any experiment. However its is clear that 
the explanation of the origin has to come of somewhere beyond SM and GUTs. In the 
last 15 years we have learned that string theory has the necessary ingredients to
solve these potential problems and it is a serious candidate to provide us with a complete unified 
theory of all known fundamental interactions of nature. In these lectures we 
attempt to give the very basic notions of some topics of perturbative and non-perturbative
string theory.

\vskip 2truecm
\newsec{Perturbative String and Superstring Theories}

In this section we overview some basic aspects of bosonic and fermionic
strings. We focus mainly in the description of the spectrum of the theory
in the light-cone gauge, the effective action, the description of spectra of
the five consistent superstring theories and the perturbative Calabi-Yau 
compactifications (for details, precisions and further
developments see for instance  Refs. \refs{\gsw,\polbook,\lt,\hatf,\oy,\kirone}).

First of all consider, as usual, the action of a relativistic point
particle. It is given by  $S = -m \int
d{\tau} \sqrt{- \dot{X}^{I}\dot{X}_{I}}$, where $X^{I}$ are $D$ functions
representing the coordinates of the $(D-1,1)$-dimensional Minkowski spacetime
(the target space), $\dot{X}^{I} \equiv {d X^{I} \over d \tau}$ and   
$m$ can be identified with the mass of
the point
particle. This action is
proportional
to the length of the world-line of the relativistic particle.

In analogy with the relativistic point particle, the action describing the dynamics of a
string (one-dimensional object) moving in a $(D-1,1)$-dimensional Minkowski spacetime (the
target space) is proportional to the area ${\bf A}$ of the worldsheet $\Sigma$. We know from the
theory of surfaces that such an area is given by ${\bf A}=\int \sqrt{det(-g)}$, where $g$ is
the induced metric (with signature $(-,+)$) on the worldsheet $\Sigma$. The background metric will be
denoted by $\eta_{IJ}$ and $\sigma^a=(\tau,\sigma)$ with $a=0,1$ are the local
coordinates on the worldsheet. $\eta_{IJ}$ and $g_{ab}$ are related by $g_{ab}=
\eta_{IJ} {\partial}_a X^{I} {\partial}_b X^{J}$ with $I , J =0,1, \dots , D-1$.
Thus
the classical action of a relativistic string is given by the Nambu-Goto action

\eqn\ocho{
S_{NG}[X^{I}]=- T \int_{\Sigma} d\tau d\sigma
\sqrt{-det({\partial}_a X^{I} {\partial}_b X^{J} \eta_{IJ})},}
where $T= {1\over 2 \pi {\alpha}'}$ is the string tension, $X^I$ are $D$ embedding
functions of the worldsheet $\Sigma$ into
the target space $X$. Now introduce a metric $h$ describing the intrinsic worldsheet geometry,
we get a classically  equivalent action to the Nambu-Goto action. This is the Polyakov action
originally proposed by Brink, di Vecchia, Howe and Zumino
\eqn\nueve{
S_P[X^I,h_{ab}] = - {1\over 4 \pi {\alpha}'} \int_{\Sigma} d^2 \sigma
\sqrt{-h}h^{ab}\partial_a X^I \partial_b X^J \eta_{IJ},}
where the $X^{I}$'s are $D$ scalar fields on the worldsheet. Such a fields can be interpreted
as the coordinates of spacetime $X$ (target space), $h$ = det$({h}^{ab})$ and
$h_{ab}={\partial}_aX^{I}{\partial}_bX^{J} \eta_{IJ}$.

Polyakov action has the following symmetries: $(i)$ Poincar\'e invariance,
$(ii)$ Worldsheet diffeomorphism invariance, and $(iii)$ Weyl invariance
(rescaling invariance).  The energy-momentum tensor of the two-dimensional
theory is given by

\eqn\diez{
T^{ab}:= {1 \over \sqrt{- h}} {\delta S_P \over \delta h_{ab}}
={1\over 4 \pi {\alpha}'} \bigg({\partial}^a X^{I}{\partial}^b X_{I}-
{1\over 2}{h}^{ab} h^{cd} {\partial}_c X^{I}
{\partial}_d X_{I} \bigg).}

Invariance under worldsheet diffeomorphisms implies that it should be conserved {\it
i.e.}
${\nabla}_aT^{ab}=0$, while the Weyl invariance gives the traceless condition,
$T^a_a=0$. The equation of motion associated with Polyakov action is given by
\eqn\once{
\partial_a \bigg( \sqrt{-h} h^{ab} \partial_b X^{I} \bigg) = 0.}
Whose solutions should satisfy the boundary conditions for the open string:  
${\partial}_{\sigma}X^{I} {\mid}^{\ell=\pi}_0=0$ (Neumann) and for the closed
string: $X^{I} (\tau , \sigma )=X^{I}(\tau , \sigma + 2 \pi)$ (Dirichlet). Here $\ell=\pi$ is
the
characteristic
length of the open string. The variation of $S_P$ with respect to $h^{ab}$ leads
to the constraint equations: $T_{ab} = 0$.
From now on we will work in the {\it conformal gauge}. In this gauge: $h_{ab} = \eta_{ab}$
and equations of motion (3.4) reduce to the Laplace equation in the flat worldsheet
whose solutions can be written as linear superposition of plane waves.

\vskip 3truecm

\vskip 1truecm
\noindent
{\it The Closed String}

For the closed string the boundary condition $X^{I}(\tau,  \sigma )= X^{I}(\tau , \sigma +
2\pi )$, leads to the general solution of Eq. (3.4)

\eqn\doce{
X^{I}(\tau,\sigma) = X^{I}_0 + {1 \over \pi T} P^{I}\tau + {i \over 2\sqrt{\pi T}}
\sum_{n \neq 0}
{1\over n} \bigg\{ \alpha^{I}_n exp\bigg(-i2n(\tau - \sigma
)\bigg) + \widetilde{\alpha}^{I}_n exp\bigg(-i2n(\tau + \sigma )\bigg)\bigg\}}
where $X^{I}_0$ and $P^{I}$ are the position and momentum of the
center-of-mass of the string and $\alpha^{I}_n$ and
$\widetilde{\alpha}^{I}_n$ satisfy the conditions
${\alpha}_n^{ I *}={\alpha}^{I}_{-n}$ (left-movers) and $
\widetilde{\alpha}_n^{I *}=\widetilde{\alpha}^{I}_{-n}$ (right-movers).

\vskip 1truecm
\noindent
{\it The Open String}

For the open string the corresponding boundary condition is ${\partial}_{\sigma}X^{I}
{\mid}^{\ell =
\pi}_0=0 $
(this is the only boundary condition which is Lorentz invariant) and the solution
is given by

\eqn\trece{
X^{I}(\tau , \sigma )= X^{I}_0 + {1 \over \pi T}P^{I}\tau +{i\over \sqrt{\pi T}}
\sum_{n \neq 0} {1 \over n} {\alpha}^{I}_nexp \big(-in\tau\big) \cos (n\sigma )}
with the matching condition ${\alpha}^{I}_n = \widetilde{\alpha}^{I}_{-n}.$

\vskip 1truecm
\noindent
{\it Quantization}

The quantization of the closed bosonic string can be carried over, as usual, by using the Dirac
prescription to the center-of-mass and oscillator variables in the form

$$
[X^{I}_0,P^{J}]=i{\eta}^{IJ},$$
$$
[{\alpha}^{I}_m,{\alpha}^{J}_n]=
[\widetilde{\alpha}^{I}_m,\widetilde{\alpha}^{J}_n]=m{\delta}_{m+n,0}{\eta}^{IJ},
$$
\eqn\catorce{
[{\alpha}^{I}_m,\widetilde{\alpha}^{J}_n]=0.}
One can identify $({\alpha}^{I}_n,\widetilde{\alpha}^{I}_n)$ with the annihilation
operators
and the corresponding operators $({\alpha}^{I}_{-n},\widetilde{\alpha}^{I}_{-n})$ with
the
creation ones. In order to specify the physical states we first denote the center of mass
state given by $|P^{I}\rangle$. The vacuum state is defined by $ {\alpha}^{I}_m
|0,P^{I} \rangle=0$ with $m > 0$ and $P^{I}|0,P^{I} \rangle =p^{I}\mid 0,P^{I}
\rangle$ and similar for the right movings (here $|0,P^{I} \rangle = |P^{I}\rangle
\otimes
|0 \rangle$). For the zero modes these states have negative
norm (ghosts). However one can choice a suitable gauge where ghosts decouple from the Hilbert
space when $D=26$. 


\vskip 1truecm
\noindent
{\it Light-cone Quantization}   

Now we turn out to work in the so called {\it light-cone gauge}. In this
gauge it is possible to solve explicitly the Virasoro constraints: $T_{ab}=0$. This is done
by removing the light-cone coordinates $X^{\pm} = {1\over \sqrt{2}}(X^0\pm
X^D)$ leaving only the transverse coordinates $X^i$ representing the physical
degrees of freedom (with $i,j =1, 2, \dots , D-2$). In this gauge the Virasoro constraints
are explicitly
solved. Thus the independent variables are $(X_0^-,P^+,X^j_0,P^j,
\alpha_n^j, \widetilde{\alpha}_n^j)$. Operators $\alpha_n^-$  and $\widetilde{\alpha}_n^-$
can be written in terms of $\alpha^j_n$ and $\widetilde{\alpha}^j_n$ respectively as   
follows: ${\alpha}^-_n={1\over \sqrt{2 \alpha
'}P^+}(\sum_{m= - \infty}^{\infty} :{\alpha}^i_{n-m}{\alpha}^i_m:-2A{\delta}_n)
$ and $\widetilde{\alpha}_n^- = {1\over \sqrt{2 \alpha
'}P^+}(\sum_{m= - \infty}^{\infty} :\widetilde{\alpha}^i_{n-m} 
\widetilde{\alpha}^i_m:-2A{\delta}_n$). For the
open string we get
${\alpha}^-_n={1\over 2\sqrt{2 \alpha '}P^+}(\sum_{m = -\infty}^{\infty} :
{\alpha}^i_{n-m} {\alpha}^i_m:-2A{\delta}_n)
$. Here $: \cdot : $ stands for the normal ordering and $A$ is its associated constant.

In this gauge the Hamiltonian is given by

\eqn\quince{
H={1\over 2}(P^i)^2+N-A \  {\rm  (open \ string),}
\ \ \ 
H=(P^i)^2+ N_L + N_R-2A \  {\rm (closed  \ string)}}
where $N$ is the operator number, $N_L = \sum_{m = - \infty}^{\infty} : \alpha_{-m} \alpha_m:$, and
$N_R = \sum_{m=- \infty}^{\infty}: \widetilde{\alpha}_{-m} \widetilde{\alpha}_m:.$ The
{\it mass-shell condition} is given by $ \alpha
' M^2= (N-A)$ (open string) and $\alpha ' M^2=2(N_L + N_R-2A)$ (closed string). For the open string,
Lorentz invariance implies that the first excited state is massless and therefore $A=1$. In the
light-cone
gauge $A$ takes the form $A = - {D-2 \over 2} \sum_{n=1}^{\infty} n$.
From the fact $\sum_{n=1}^{\infty}n^{-s}=\zeta(s),$ where $\zeta$
is the Riemann's zeta function (which converges for $s>1$ and has a unique analytic continuation at
$s=-1$, where it takes the value $-{1\over 12}$) then  $A=-{D-2\over
24}$ and therefore  $D=26$.

\vskip 1truecm
\subsec{Spectrum of the Bosonic String}

\noindent
{\it Closed Strings}

The spectrum of the closed string can be obtained from the combination of the left-moving
states and the right-moving ones. The ground state ($N_L =N_R =0$) is given by
$\alpha 'M^2=-4$. That means that the ground state includes a tachyon. The first
excited state ($N_L=1=N_R$) is massless and it is given by
${\alpha}^{i}_{-1} \widetilde{\alpha}^{j}_{-1} |0,P \rangle$. This state can be naturally
decomposed into irreducible representations of the little group $SO(24)$ as follows

$$
{\alpha}^i_{-1}\widetilde{\alpha}^j_{-1}\mid 0,P \rangle=
{\alpha}^{[i}_{-1}\widetilde{\alpha}^{j]}_{-1}\mid
0,P \rangle + \bigg({\alpha}^{(i}_{-1}\widetilde{\alpha}^{j)}_{-1}-{1\over D-2}{\delta}^{ij}
{\alpha}^k_{-1}\widetilde{\alpha}^k_{-1}\bigg)\mid
0,P \rangle$$
\eqn\dseis{
+~{1\over D-2}{\delta}^{ij}{\alpha}^k_{-1}\widetilde{\alpha}^k_{-1}\mid 0,P \rangle .}
The first term of the rhs is interpreted as a spin 2 massless particle
$G_{ij}$ ({\it graviton}).  The second term is a range 2 anti-symmetric tensor
$B_{ij}$. While the last term is an scalar field
$\Phi$ ({\it dilaton}). Higher excited massive states are combinations of 
irreducible representations of the corresponding little group SO$(25).$

\vskip 1truecm

\noindent
{\it Open Strings}

For the open string, the ground state includes once again a tachyon since
$\alpha 'M^2=-1$. The first exited state
$N=1$ is given by a massless vector field in 26 dimensions. The second excitation level
is given by the massive states ${\alpha}^{i}_{-2}\mid 0,P \rangle$ and
${\alpha}^{i}_{-1}{\alpha}^{j}_{-1}\mid 0,P \rangle$ which are in irreducible
representations of the little group SO$(25)$.

\vskip 1truecm
\subsec{Interacting Strings and the Effective Action}

\noindent
{\it Interacting Strings}

So far we have described the {\it free} propagation of a closed (or open) bosonic string. In what
follows we consider the interaction of these strings. Here we focus in the closed string case,
the open case requires from further definitions. The interaction of strings at the perturbative
level is just the extension of the technique of Feynman diagrams for point particles to extended
objets. The vacuum-vacuum amplitude ${\cal A}$ is given by 

\eqn\dsiete{
{\cal A} \sim \int {\cal D} h_{ab} {\cal D} X^I exp \bigg(i S_P[X^I,h_{ab}] \bigg).}

The interacting case requires to sum  over all loop diagrams. In the closed string case it means
that we have to sum over all compact orientable surfaces with non-trivial boundary
($\partial \Sigma \not= 0$). In two
dimensions these surfaces are
completely characterized by their number of holes $g$ (the genus) and boundaries $b$. The
relevant topological invariant is the Euler number $\chi(\Sigma) = {1 \over 4 \pi} \int_{\Sigma} d^2
\sigma \sqrt{-h}
R^{(2)}$, where $R^{(2)}$ is the scalar curvature of the worldsheet $\Sigma$. In order to
include the interaction of strings, the generalization of the Polyakov action consistent with 
its symmetries is given by

\eqn\docho{
S= S_P[X^I,h_{ab}] + {\Phi(X^I) \over 4 \pi} \int_{\Sigma} d^2 \sigma  \sqrt{-h} R^{(2)}
+ {1 \over 2 \pi} \int_{\partial \Sigma} ds K,}
where $\Phi(X)$ is an scalar background field and represents the
gravitational coupling constant of the two-dimensional Einstein-Hilbert Lagrangian.  
$K$ in the above equation stands for the geodesic curvature of $\Sigma$. 
If we define the string coupling constant by $g_S \equiv e^{\Phi},$ then Eq. (3.10)
for the case of closed strings generalizes to

\eqn\dnueve{
{\cal A} \sim \sum_{\chi} g_S^{\chi(\Sigma)} \int {\cal D} h_{ab} {\cal D} X^I 
exp \bigg(i S_P[X^I,h_{ab}]
\bigg).}
The amplitude defined on-shell correspond to $g=0$ and the rest ($g\geq 1$) corresponds to
$g$-loop corrections. 

The definition of correlation functions of operators requires of the idea of {\it vertex
operators} ${\cal W}_{\Lambda}$. These operators are defined as

\eqn\veinte{
{\cal V}_{\Lambda}(k) = \int d^2 \sigma \sqrt{-h} {\cal W}_{\Lambda}(\sigma,\tau) exp\big(
ik\cdot X),}
where ${\cal W}_{\Lambda}(\sigma,\tau)$ (with $\Lambda$ being a generic massless field of the bosonic
spectra) is a local operator assigned
to some specific state of the
spectrum of the theory. For instance for the tachyon it is given by ${\cal
W}_{T}(\sigma,\tau)
\sim \partial_a X_{I} \partial^a X^{I}$. While that for the graviton $G$ with polarization
$\zeta_{IJ}$  it is given by
$W_{G}(\sigma,\tau) = \zeta_{IJ}\partial_a X^{I} \partial^a X^{J}.$  Local operators  
${\cal V}_{\Lambda}$ are diffeomorphism and conformal invariant and therefore more convenient 
to define scattering amplitudes. 

Thus one can define the scattering amplitude of the vertex field operators by
their 
corresponding invariant operators ${\cal V}_{\Lambda}$. In perturbation theory the
scattering amplitude is given by

\eqn\vuno{
{\cal A}(\Lambda_1,k_1; \dots \Lambda_N,k_N) \sim \sum_{\chi} g_S^{\chi(\Sigma)}
\int {\cal D} h {\cal D} X exp \bigg(iS_P[X^I,h_{ab}] \bigg) \prod_{i=1}^N {\cal
V}_{\Lambda_i}(k_i).}
This scattering amplitude is, of course, proportional to the correlation function of the 
product of $N$ invariant operators ${\cal V}_{\Lambda_i}(k_i)$ as follows

\eqn\vdos{
{\cal A}(\Lambda_1,k_1; \dots \Lambda_N,k_N) \propto \langle \prod_{i=1}^N {\cal
V}_{\Lambda_i}(k_i) \rangle .}

\vskip 3truecm

\vskip 1truecm
\noindent
{\it Effective String Actions}

In order to make contact with the spacetime physics we now describe how the
spacetime equations of motion come from conformal invariance conditions for the
non-linear sigma model in curved spaces. The immediate generalization of the Polyakov
action is 

\eqn\vtres{
S = - {1 \over 4 \pi \alpha '} \int_{\Sigma} d^2 \sigma \sqrt{-h} h^{ab} \partial_a X^{I}
\partial_b
X^{J} G_{IJ}(X),}
where $G_{IJ}(X)$ is an arbitrary background metric of the curved target space $X$. The
perturbation of this metric $G_{IJ}(X) = \eta_{IJ} + h_{IJ}(X)$
in the partition function  $Z\sim exp \big(-S[X^I, \eta_{IJ} + h_{IJ}]\big)$, leads to
an expansion in powers of $h_{IJ}.$  This partition
function can be easily interpreted as containing the information of the interaction of the
string with a coherent state of gravitons with invariant operator ${\cal V}_G(k) = \int
d^2 \sigma  \sqrt{-h} {\cal W}_{G}(\sigma,\tau) exp \big( ik\cdot X \big)$ with
${\cal W}_G =  h^{ab} \partial_a X^{I} \partial_b X^{J} h_{IJ}(X)$. 

On the other hand, the Polyakov action can be generalized to be consistent with 
all symmetries and  with the massless spectrum of the
bosonic closed strings in the form of a non-linear sigma model

\eqn\vcuatro{
\widehat{S}= {1 \over 4 \pi \alpha '} \int_{\Sigma} d^2 \sigma \sqrt{-h} \bigg[
\bigg( h^{ab} G_{IJ}(X) +i \varepsilon^{ab} B_{IJ}(X) \bigg) \partial_a X^{I}
\partial_b X^{J} + \alpha ' \Phi(X) R^{(2)} \bigg],}
where $G_{IJ}(X)$ is the target space curved {\it metric}, $B_{IJ}(X)$ is an
anti-symmetric  field, also called  the {\it Kalb-Ramond} field, and $\Phi(X)$ is the scalar
field called the {\it dilaton} field. From the viewpoint of the two-dimensional
non-linear sigma model these {\it background} fields can be regarded as {\it coupling 
constants} and the renormalization group techniques become applied. The computation of the
quantum {\it conformal anomaly} by using the dimensional regularization method, leads to
express the energy-momentum trace as a linear combination 

\eqn\vcinco{
T^a_a = -{1 \over 2 \alpha '} \beta^G_{IJ} h^{ab} \partial_a X^{I}
\partial_b X^{J}  -  {i \over 2 \alpha '} \beta^B_{IJ}\varepsilon^{ab} 
\partial_a X^{I} \partial_b X^{J} - {1\over 2} \beta^{\Phi} R^{(2)},}
where $\beta$ are the one-loop beta functions associated with each coupling 
constant or background field. They are explicitly computed and give

\eqn\vseis{
\beta^G_{IJ} = \alpha ' \bigg( R_{IJ} + 2 \nabla_{I}  \nabla_{J}
\Phi -{1 \over 4} H_{IKL} H^{KL}_{J} \bigg) + O(\alpha'^2),}

\eqn\vsiete{
\beta^B_{IJ} = \alpha ' \bigg(  - {1\over 2} \nabla^{K}  H_{KIJ}
+ \nabla^{K}\Phi H_{KIJ} \bigg) + O(\alpha'^2),}

\eqn\vocho{
\beta^{\Phi} = \alpha ' \bigg( {D-26 \over 6 \alpha '} -{1 \over 2} \nabla^2 \Phi
+ \nabla_{K} \Phi \nabla^{K} \Phi -{1 \over 24} H_{IJK} H^{IJK} \bigg) + O(\alpha'^2),}
where $H_{IJK} = \partial_{I} B_{J K} + \partial_{J} B_{KI}
+ \partial_{K} B_{IJ}.$
Weyl invariance at the quantum level implies the vanishing of the conformal anomaly
and therefore the vanishing of each beta function. This leads to three coupled
field equations for the background fields. These conditions for these fields can been
regarded as equations of motion derivable from the spacetime action in $D$ dimensions

\eqn\vnueve{
S={1 \over 2 \kappa^2_0} \int_X d^Dx \sqrt{-G} e^{-2 \Phi} \bigg(
R + 4 \nabla_{I}\Phi \nabla^{I}\Phi - {1 \over 12} H_{IJK} H^{IJK} - {2(D-26) \over 3 \alpha
'} 
+ O(\alpha ') \bigg),}
where $\kappa_0$ is a normalization constant.

It is interesting to see that a redefinition of background metric under the
transformation in $D$ dimensions  $\widetilde{G}_{IJ}(X) = exp\big(2 \varpi(X)\big)
G_{IJ}$
with $\varpi = {2 \over D-2} (\Phi_0 - \Phi)$ leads to the background action in the
`Einstein frame'

$$
\widehat{S}={1 \over 2 \kappa^2} \int d^Dx \sqrt{-\widetilde{G}} \bigg(
\widetilde{R} - {4 \over D-2}\nabla_{I} \widetilde{\Phi} \nabla^{I}\widetilde{\Phi} 
- {1 \over 12} e^{-8 \widetilde{\Phi} / (D-2)} H_{IJK} H^{IJK}
$$
\eqn\treinta{
- {2(D-26) \over 3 \alpha '} e^{ 4 \widetilde{\Phi} / (D-2)} + O(\alpha ') \bigg),}
where $\widetilde{R} = e^{-2 \varpi} \big[ R - 2 (D-1) \nabla^2 \varpi -(D-2)(D-1) 
\partial_{I} \varpi \partial^{I} \varpi \big]$ and $\widetilde{\Phi} = \Phi - \Phi_0.$
The form of this action will of extreme importance later when we describe the strong/weak
coupling duality in effective supergravity actions of the different superstring theories
types. In the above action $\kappa \equiv \kappa_0 e^{\Phi_0}= \kappa_0 \cdot g_S$ is the
gravitational coupling constant in $D$ dimensions, {\it i.e.} $\kappa = \sqrt{8 \pi G_N}.$

A very close procedure can be performed for the open string and compute its effective low
energy action. It was done about 15 years ago and for  gauge fields with constant
field strength $F_{IJ}$ it is given by the Dirac-Born-Infeld action

\eqn\tuno{
S_O= - T \int d^Dx e^{- \Phi} \sqrt{ - det \big( G_{IJ} +  B_{IJ} +
2 \pi \alpha ' F_{IJ} \big)}.}
Later we shall talk about some applications of this effective action.

\vskip 1truecm
\subsec{Superstrings}

In bosonic string theory there are two bold problems. The first one is the presence
of tachyons in the spectrum. The second one is that there are no spacetime fermions.
Here is where superstrings come to the rescue. A superstring is described, despite of
the usual bosonic fields $X^{I}$, by
fermionic fields $\psi^{I} _{L,R}$ on the worldsheet $\Sigma$. Which satisfy anticommutation rules and
where
the $L$ and $R$ denote the left and right worldsheet chirality respectively. The action for the
superstring is given by

\eqn\tdos{
L_{SS}=-{1\over 8\pi}\int d^2 \sigma
\sqrt{-h} \bigg(h^{ab}\partial_a X^I \partial_b X_I +
2i \bar{\psi}^{I}\gamma^{a}\partial_{a}\psi_{I} -i \bar{\chi}_a\gamma^b
\gamma^a\psi^I \big(\partial_bX_I-{i\over 4}\bar{\chi}_b\psi_I \big)\bigg),}
where  $\psi^{I}$ and $\chi_a$ are the superpartners of $X^{I}$ and the tetrad field $e^a$
respectively.
In the superconformal gauge ($h_{ab} = \eta_{ab}$ and $\chi_a = 0$) and in light-cone
coordinates it can be reduced to

\eqn\ttres{
L_{SS}= {1 \over 2 \pi}
\int \bigg(\partial_L X^{I}  \partial_R
X_{I} + i \psi^{I}_R\partial_L \psi_{I\  R} +
i \psi^{I}_L \partial_R \psi_{I \ L} \bigg).}

In analogy to the bosonic case, the local dynamics
of the worldsheet metric is manifestly conformal anomaly free at the quantum level if 
the  critical
spacetime dimension $D$ is 10. Thus the string
oscillates in the 8 transverse dimensions. The action (32) is
invariant under: $(i)$ worldsheet supersymmetry, $(ii)$ Weyl transformations,
$(iii)$ super-Weyl transformations, $(iv)$ Poincar\'e transformations and
$(v)$ Worldsheet reparametrizations. The equation of motion for the $X's$ fields is
the same that in the bosonic case (Laplace equation) and
whose general solution is given by Eqs. (3.5) or (3.6). Equation of
motion for the fermionic field is the Dirac equation in two dimensions.
Constraints here are more involved and they are called the {\it super-Virasoro
constraints}. However in the light-cone gauge, everything simplifies and
the transverse coordinates (eight coordinates) become the bosonic physical degrees
of freedom together with their corresponding supersymmetric partners. Analogously
to the bosonic case,
massless states of the spectrum come into representations of the little group SO(8)
which is a subgroup of SO$(9,1)$, while that the massive states lie into representations of the little
group SO$(9)$.

For the closed string there are two possibilities for the boundary conditions of fermions:
$(i)$ periodic
boundary conditions (Ramond ({\bf R}) sector)
$\psi^{I}_{L,R}(\sigma) = + \psi^{I}_{L,R}(\sigma + 2 \pi)$
and $(ii)$
anti-periodic boundary conditions
(Neveu-Schwarz ({\bf NS}) sector) $\psi^{I}_{L,R}(\sigma) = - \psi^{I}_{L,R}(\sigma + 2 \pi)$.
Solutions of Dirac equation satisfying these boundary conditions are

\eqn\tcuatro{
\psi^{I}_L(\sigma,\tau) = \sum_n \bar{\psi}^{I}_{-n}exp\bigg(-in(\tau +\sigma  
)\bigg), \ \ \ \ \  \psi^{I}_R(\sigma,\tau) =
\sum_n \psi^{I}_{n}exp \bigg(-in(\tau -\sigma
)\bigg),}
where $\bar{\psi}^{I}_{-n}$ and $\psi^{I}_{n}$ are fermionic modes of left and right movers
respectively. 

In the case
of the fermions in the {\bf R} sector
$n$ is integer and it is semi-integer in the {\bf NS} sector.

The quantization of the
superstring come from the promotion of the fields $X^{I}$ and $\psi^{I}$ to operators
whose oscillator variables are operators satisfying the relations
$[\alpha^I _n, \alpha^J _m]_-~= n\delta_{m+n,0}\eta^{IJ}$ and
$[\psi^I _n,\psi^J _m]_+~=~\eta^{IJ}\delta_{m+n,0},$ where $[,]_-$ and
$[,]_+$ stand for commutator and anti-commutator respectively.

The zero modes of $\alpha$ are diagonal in the Fock space and its
eigenvalue can be identified with its momentum. For the {\bf NS} sector there is no
fermionic zero modes but they can exist for the {\bf R} sector and they  satisfy a Clifford
algebra $[\psi^I _0,\psi^J _0]_+~=~\eta^{IJ}$. The Hamiltonian for the
closed superstring is given by $H_{L,R}=N_{L,R}+ {1\over 2}P^2_{L,R}-A_{L,R}$. For the
{\bf NS} sector $A={1 \over 2}$, while for the {\bf R} sector $A=0$. The mass is given by
$M^2 = M_L^2 + M_R^2$ with
${1\over 2} M^2_{L,R}=N_{L,R}-A_{L,R}$.

There are five consistent  superstring theories: Type IIA, IIB, Type I, SO(32) and
$E_8 \times E_8$ heterotic strings, represented by HO and HE respectively. In what follows of this
section we briefly describe the spectrum
in
each one of them.

\vskip 1truecm
\noindent
{\it Type II Superstring Theories}

In this case the theory consist of closed strings only. They are theories with ${\cal N} =2$
spacetime supersymmetry. There are 8 scalar fields (representing
the 8 transverse coordinates to the string) and one Weyl-Majorana spinor. There are
8 left-moving and 8 right-moving fermions.

In the {\bf NS} sector there is still a tachyon in the ground state.  But in the
supersymmetric case this problem can be solved through the introduction of
the called {\bf GSO} projection. This projection eliminates the tachyon in the {\bf NS} sector
and it acts in the {\bf R} sector as a ten-dimensional spacetime chirality operator. That means
that the
application of the {\bf GSO} projection operator defines the chirality of a
massless spinor in the {\bf R} sector. Thus from the left and right
moving sectors, one can construct states in
four different sectors: $(i)$ {\bf NS-NS},
$(ii)$ {\bf NS-R}, $(iii)$ {\bf R-NS} and $(iv)$ {\bf R-R}. Taking into account the two types of
chirality $L$ and $R$ one has two possibilities:

\noindent
$a)-$ The {\bf GSO} projections on the left and right fermions produce different
chirality in the ground state of the {\bf R} sector ({\it Type IIA}).

\noindent
$b)-${\bf GSO} projection are equal in left and right sectors and the ground states
in the {\bf R} sector, have the same chirality ({\it Type IIB}). Thus the spectrum for the Type IIA
and
IIB superstring theories is:

\item{$\bullet$} {\it Type IIA}

The {\bf NS-NS} sector has a symmetric tensor field $G_{IJ}$ (spacetime metric), an
antisymmetric
tensor field
$B_{IJ}$ and a scalar field $\Phi$ (dilaton). In the {\bf R-R} sector there is a vector field
$A_{I}$ associated with a 1-form $A_{(1)}$ ($A_{I} \Leftrightarrow A_{(1)}$) and a rank 3
totally
antisymmetric tensor $A_{IJK} \Leftrightarrow A_{(3)}$ and by Hodge duality in ten
dimensions also we have $A_{(5)}$, $A_{(7)}$ and $A_{(9)}$. In general the {\bf R-R} sector
consist of $p$-forms $F_{(p)}= dA_{(p-1)}$ (where $A_{(p)}$ are called RR fields) on the
ten-dimensional spacetime $X$ with $p$ {\it even} {\it i.e.}
$F_{(2)}, F_{(4)},  \dots , F_{(10)}$. In the {\bf NS-R} and {\bf R-NS} sectors we have
two gravitinos with opposite
chirality and the supersymmetric partners of the mentioned bosonic fields.

\item{$\bullet$} {\it Type IIB}

In the {\bf NS-NS} sector Type IIB theory has exactly the same spectrum that of Type IIA theory.
On the {\bf R-R} sector it has a scalar field $a \Leftrightarrow A_{(0)}$ (the axion field), an
antisymmetric tensor
field $B'_{IJ} \Leftrightarrow A_{(2)}$ and a rank 4 totally antisymmetric tensor $D_{IJKL}
\Leftrightarrow A_{(4)}$
whose field strength is self-dual {\it i.e.},
$ F_{(5)}= d A_{(4)}$ with $*F_{(5)} = + F_{(5)}$. Similar than for the case of Type IIA
theory one has also the Hodge dual fields $A_{(6)},$ $A_{(8)},$ $A_{(10)}$.
In general,
RR fields in Type IIB theory are given by $p$-forms $F_{(p)} = d A_{(p-1)}$ on the spacetime
$X$ with $p$ {\it odd}
{\it i.e.}
$F_{(1)}, F_{(3)}, \dots , F_{(11)}$. The {\bf NS-R} and {\bf R-NS} sectors do contain two
gravitinos
with the
same chirality and the corresponding fermionic matter.

\vskip 1truecm
\subsec{Type I Superstrings}

In this case the $L$ and $R$ degrees of freedom are identified. Type I and Type IIB theories
have the
same spectrum, except that in the former one the states which are not invariant
under the change of orientation of the worldsheet, are projected out. This worldsheet parity
$\Omega$ interchanges
the left and right modes. Type I superstring theory is a theory of breakable closed strings, thus it
incorporates also open strings. The $\Omega$ operation leave invariant only one half of
the spacetime supersymmetry, thus the theory is ${\cal N}=1$.

The spectrum of bosonic massless states in the {\bf NS-NS} sector is:
$G_{IJ}$ (spacetime metric) and $\Phi$ (dilaton) from the closed sector and $B_{IJ}$
is
projected
out.  On the {\bf R-R} sector there is an
antisymmetric field $B_{IJ}$ of the closed sector. The open string sector is necessary in
order to cancel tadpole diagrams. A contribution to the spectrum come from this sector. Chan-Paton
factors can be added at the boundaries of open strings. Hence the cancellation of the tadpole are
needed 32 labels at each end. Therefore in the {\bf NS-NS} sector there are 496 gauge fields in the
adjoint representation of SO(32).

\vskip 1truecm
\subsec{Heterotic Superstrings}

This kind of theory involves only closed strings. Thus there are left and right sectors. The
left-moving sector contains a bosonic string theory and the right-moving sector contains
superstrings. This theory is supersymmetric on the right sector only, thus the theory contains
${\cal N}=1$ spacetime supersymmetry. The momentum at the left sector $P_L$ lives in 26 dimensions,
while
$P_R$
lives in 10 dimensions. It is natural to identify the first ten components
of $P_L$ with $P_R$. Consistency of the theory tell us that the extra 16 dimensions should
belong to the root lattice $E_8 \times E_8$ or a $\IZ_2$-sublattice of the SO(32) weight
lattice.

The spectrum consists of a tachyon in the ground state of the left-moving sector. In both sectors
we have the spacetime metric $G_{IJ}$, the antisymmetric tensor $B_{IJ}$, the dilaton
$\phi$ and finally there are 496 gauge fields $A_I$ in the adjoint representation
of the gauge group $E_8 \times E_8$ or SO$(32)$.

All these Types of superstring theories do admit a low energy effective description in terms 
of a supergravity theory. These theories involves the corresponding background fields of their
spectra. Supergravity actions of these diverse types will be constructed later to study strong/weak
coupling duality in string theory.

\vskip 1truecm
\subsec{Calabi-Yau Compactifications in Perturbative String Theory}

In order to connect superstring theories to the observed 4-dimensional spacetime
physics, 
we have to reduce  the critical dimension $D=10$ to four dimensions. To preserve 
certain supersymmetry consistent with chirality in four dimensions 
it is necessary to require some properties to the ten
dimensional spacetime $X$. Perhaps the simplest ansatz is to assume that 
the four-dimensional Minkowski spacetime $M$ and a six-dimensional internal space ${\cal
K}$ {\it factorizes}
as $X \cong M \times {\cal K}$,  where ${\cal K}$ has tiny dimensions and
unobservable in our present experiments. It is worth to say that this factorization ansatz is not
unique and other
possibility is the {\it warped compactification} of the celebrated Randall-Sundrum
scenarios, which are nicely reviewed in Ref. \shamit. 

It is useful to classify the compactifications according to how much 
supersymmetries
is broken, because this number is related with the quantum corrections that  
we shall consider.  We choose ${\cal K}$ to be a
manifold with the property that a 
certain
number of supersymmetries are preserved\foot{Cosmological constant is
generated by perturbation theory. Strings propagating in a ${\cal K}$ manifold in 
which all
supersymmetries are broken distablizes the Minkowski vacuum.}.

We are now looking for conditions in the background which leave some 
supersymmetry
unbroken. These conditions are given by null variations of the Fermi fields.

Consider the {\it diagonal} metric for ten-dimensional spacetime $X$ given by $G_{IJ}= f(y) \eta_{\mu
\nu} +
G_{mn}(y)$ where $y$ denotes the compactified coordinates and $I,J=0,...,9$, $\mu ,\nu
=0,...,3$, $m,n=4,...,9$. For $D=10$, ${\cal N}=1$ heterotic string theory the Fermi fields
variations are:

\item{$\bullet$} {\it gravitino}:  $\nabla \psi_{\mu}=\Delta_{\mu}\varepsilon$,  
    $\delta \psi_m=(\partial_m+{1\over 4}\Omega^-_{mnp}\Gamma^{np})\varepsilon,$

\item{$\bullet$} {\it dilatino}:          $\delta \xi =(\Gamma^m\partial_m\phi
-{1\over 12}\Gamma^{mnp}H_{mnp})\varepsilon,$

\item{$\bullet$}{\it gaugino}:      $\delta \lambda = F_{mn}\Gamma^{mn}\varepsilon,$

\noindent
where $\varepsilon$ is a Weyl-Majorana spinor in ten dimensions, $\Omega^-_{mnp}$ is the
internal component of $\Omega^-_{MNP} = \omega_{MNP} - {1 \over 2} H_{MNP}$, $\Gamma$ are the
Dirac matrices.

The compactification ansatz $X=M \times {\cal K}$ breaks the Lorentz group SO(9,1) into
SO$(3,1) \times {\rm SO}(6)$. In the spinor representation {\bf 16}
the Weyl-Majorana supersymmetry parameter $\varepsilon_{\alpha \beta}$ 
decomposes as $\varepsilon(y) \rightarrow \varepsilon_{\alpha \beta}(y) 
+ \varepsilon^*_{\alpha \beta}(y)$ under ${\bf 16} \to ({\bf 2},{\bf 4}) \oplus 
({\bf 2}^*,{\bf 4}^*)$. The general form of $\varepsilon_{\alpha \beta}$ is 
$\varepsilon_{\alpha \beta} = u_{\alpha} \zeta_{\beta}(y)$ with $u_{\alpha}$ an
arbitrary Weyl spinor. When we put the condition that Fermi fields variations vanish, then
each internal spinor $\zeta_{\beta}(y)$ gives the  minimal (${\cal N}=1$) $D=4$ 
supersymmetric algebra.

Now, by the null Fermi fields variations we can find conditions in the background
fields assuming that $H_{mnp} =0$. These are:

\item{$\bullet$} $\delta \zeta =0 \Rightarrow \partial_m\Phi =0,$

\item{$\bullet$} $\delta \psi =0 \Longrightarrow G_{\mu \nu}=\eta_{\mu \nu},$

\item{$\bullet$} $\delta \psi_m =0 \Rightarrow \nabla_m\zeta =0.$

The last equation tell us that $\zeta_{\beta}$ is covariantly
constant on the internal space ${\cal K}$, and implies that ${\cal K}$ is Ricci-flat.
This is because
$[\nabla_m,\nabla_n]\zeta ={1\over 4}R_{mnpq}\Gamma^{pq}\zeta=0$. For this 
reason, in general
$\Gamma^{pq}$ do not belong to SO(6) but to SU(3), which is a subgroup that 
leaves one component of the spinor $\zeta$ invariant. Thus the 
compact
manifold ${\cal K}$ must have SU(3) holonomy. The second unbroken susy condition
implies that the warped factor $f(y)$ in metric is 1 and the metric $G_{IJ}$ is unwarped.
Finally, the first condition implies that the dilation is constant. 
This is a {\it Calabi-Yau} three-fold. A Calabi-Yau three-fold is also a K\"{a}hler manifold
in which the first Chern 
class zero {\it i.e.} $c_1(T{\cal K})=0.$ Any Calabi-Yau manifold possesses a unique Ricci-flat metric.
When we
consider ${\cal N}=1$ heterotic string theory  on Calabi-Yau three-fold we obtain a
four-dimensional chiral theory with spacetime supersymmetry ${\cal N}=1$.
In fact, compactification on manifolds of SU(3) holonomy preserves $1/4$ of
supersymmetry. If we consider ${\cal N}=2$ theories (for example, type II 
superstrings) in
$D=10$ dimensions, after compactification on a Calabi-Yau three-fold we obtain ${\cal N}=2$
theories in $D=4$.

In addition to the CY-threefold structure for ${\cal K}$ the unbroken susy condition 
$\delta \lambda^a = 0 = F^{a}_{mn} \Gamma^{mn} \varepsilon,$ leads to the equations in complex 
coordinates

\eqn\tcinco{
F_{IJ} = F_{\bar{I},\bar{J}}= 0, \ \ \ \ \ \  G^{I \bar{J}} F_{I \bar{J}} = 0.}
These equations require to specify a gauge subbundle $V$ of a $E_8 \times E_8$ gauge bundle
over ${\cal K}$ and a gauge connection $A$ on $V$ with curvature $F$. The condition 
$F_{IJ} = F_{\bar{I},\bar{J}}= 0$ tell us that the subbundle $V$ as well as the
corresponding connection should be holomorphic. The second condition  $G^{I\bar{J}}
F_{I \bar{J}}=0$ is the celebrated Donaldson-Uhlenbeck-Yau equation for $A$. This equation has a
unique solution if the bundle $V$ is stable and if it is satisfied the integrability condition
$\int_{\cal K} \Omega^{n-1} \wedge c_1(V) =0,$ where $\Omega$ is the K\"ahler form of
${\cal K}$. There is a further condition to be satisfied by the connection $A$, the Bianchi
identity for $H$ and $F$, it is given  by

\eqn\tseis{
dH = tr R\wedge R - {1 \over 30} tr F\wedge F.}
The only solution is $tr R\wedge R \propto  tr F\wedge F$ which implies that $c_2(T{\cal K})
= c_2(V)$. This situation is usually known as the {\it standard embedding} of the spin connection 
in the gauge connection ant it is a method to determine the connection $A$ on $V$. 

Thus in the compactification of phenomenological interest of the heterotic theory with the
ansatz $X = M \times {\cal K}$, the internal space has to be a Calabi-Yau three-fold and one
has to specify a stable, holomorphic vector bundle $V$ over $X$ (or ${\cal K}$) satisfying
$c_1(V)=0$ and $c_2(V) = c_2(TX)$. 

If $V$ is a SU$(n)$ vector bundle over $X$ the subgroups of $E_8 \times E_8$ that
commutes are $E_6$, SO(10) and SU(5) for $n=3,4,5$ respectively. This leads to 
GUTs in four dimensions justly with the gauge groups $E_6$, SO(10) or SU(5).

\vskip 1truecm
\noindent
{\it Massless Spectrum}

In order to describe the impact of the characteristics of ${\cal K}$ and $V$ on the 
properties of the spectrum of the four dimensional theory we start by decomposing
the ten-dimensional Dirac operator under $M \times {\cal K}$ into  

\eqn\tsiete{
\not \! \! D^{(10)} = \sum_{I=0}^{9} \Gamma^I D_{I} = \not \! \! D^{(4)} + \not \! \!
D_{\cal K},}
where $\not \! \! D^{(4)} = \sum_{I=0}^3 \Gamma^I D_I$ and $\not \! \! D_{\cal K} =
\sum_{J=4}^{9} \Gamma^J
D_J$. Dirac equation in ten dimensions is

\eqn\tocho{
\not \! \! D^{(10)} \Psi(x^I,y^J) = \big(\not \! \! D^{(4)} + \not \! \! D_{\cal K}
\big)
\Psi(x^I,y^J).}

Thus the spectrum of the Dirac operator $\not \! \! D_{\cal K}$ on ${\cal K}$ determines 
the massive spectrum of fermions in four dimensions.

In ten dimensions the Lorentz group only has {\it real} spinor representations and
the Clifford modules decomposes as:
$S^{(10)} = S^{(10)}_+ \oplus S^{(10)}_-$. Positive and negative chirality are distinguised by
$\Gamma^{(10)} =
\Gamma^0 \Gamma^1 \dots \Gamma^{9}$. $CPT$ theorem implies that we must take 
only one chirality

\eqn\tnueve{
\Gamma^{(10)} \Psi = + \Psi .}

Decompose the spinor representation of SO(1,9) under ${\rm SO}(1,3) \times {\rm SO}(6)$ with
$\Gamma^{(10)} = \Gamma^{(4)} \cdot \Gamma^{(6)}$ where $\Gamma^{(4)}= i \Gamma^0
\Gamma^1 \Gamma^2 \Gamma^3$ and $\Gamma^{(6)} = -i \Gamma^4 \Gamma^5 \dots \Gamma^{9}$.
One solution with $\Gamma^{(10)} = +1$ is given by $\Gamma^{(4)} = \Gamma^{(6)}$ and then the spin
bundle decomposes
under $M \times {\cal K}$ as

\eqn\cuarenta{
\widehat{S}^{(10)} = \bigg(\widehat{S}^{(4)}_+ \otimes \widehat{S}^{\cal K}_+\bigg) \oplus 
\bigg(\widehat{S}^{(4)}_- \otimes \widehat{S}^{\cal K}_-\bigg).}

Now solve the Dirac equation with the ansatz $\Psi(x^I,y^J) = \sum_m \phi_m(x^I) \otimes
\chi_m(y^J) = \sum_m \psi_m$ and $\not \! \! D'_{\cal K} \chi_m = \lambda_m \chi_m$. It leads to

\eqn\cuno{
\big(\not \! \! D'^{(4)} + \lambda_m \big) \psi_m = 0,}
where $\not \! \! D'^{(4)} = \Gamma^{(4)} \not \! \! D^{(4)}$
and $\not \! \! D'_{\cal K} = \Gamma^{(4)} \not \! \! D_{\cal K}$. 

$\not \! D_{\cal K}$ is an
elliptic operator on the 
compact manifold ${\cal K}$, this implies that that operator has a {\it finite} number of 
fermion zero modes. Massless fermions in four dimensions originate as zero
modes of the Dirac operators  $\not \! D^{\cal K}$ of the internal manifold ${\cal K}$.  By
the
Atiyah-Singer theorem, a topological invariant of ${\cal K}$ containing the information of
the chiral fermions on ${\cal K}$ is given by the index of the Dirac operator

\eqn\cdos{
Index(\not \! \! D_{\cal K}) = N^{\lambda =0}_+ - N^{\lambda =0}_-,}
for chiral fermions on ${\cal K}$ with $\Gamma^{(6)} = \pm 1.$ Here $ N^{\lambda =0}_{\pm}$
are the number of positive or negative chiral zero modes.  In $2k +2$ dimensions
this index is vanishing. We need to couple gauge fields coming from the heterotic string
theory. Recall that they are $E_8 \times E_8$ valued gauge fields. 

The {\it standard embedding} of the
spin connection in the gauge connection leads to the chain of maximal subgroups: 
$SO(6) \times SO(10) \subset
SO(16) \subset E_8$. This breaks SO(16) to SO(10). The computation of the $\Delta$
for this case yields $\Delta = \oplus_i L_i \otimes {\cal R}_i$ where $L_i$ are irreps of
SO(6) and ${\cal R}_i$ are complex irreps of SO(10). These latter determine the irreps where are
distributed the massless fermions of the  four-dimensional theory. The former irreps $L_i$ determine
the number of fermionic chiral zero modes described by the topological index  
$\delta = Index(D_{\cal K})$. This is given by

$$
\delta = N^{\lambda=0}_{\Gamma^{(6)}= +1} -  N^{\lambda=0}_{\Gamma^{(6)}= -1}
= \int_{\cal K} ch(V) td({\cal K}) = {1\over 2} \int_{\cal K} c_3(V)
$$
and from the solution $\Gamma^{(6)} = \Gamma^{(4)}$ it determines the chiral fermion families in four 
dimensions

\eqn\ctres{
\delta= N^{\lambda=0}_{\Gamma^{(4)}= +1}  -  N^{\lambda=0}_{\Gamma^{(4)}= -1}.} 
Thus the theory in four dimensions has $\Delta = \oplus_i \delta_i {\cal R}_i$ where

\eqn\ccuatro{
\Delta = \delta \bigg( {\bf 16} \ominus {\bf 16}^* \bigg),}
where  $ \delta = \chi({\cal K})/2$ with $\chi({\cal K})$ is the Euler number of ${\cal K}.$ 

\vskip 1truecm
\subsec{Some Physics in Four Dimensions}

In this section we intend to make contact with some four-dimensional physics.
The development of this line of work is known as {\it string phenomenology}.
Recent reviews of this topic at the light of string dualities is given in 
Ref. \ibanez. In the present short review we follows Refs. \refs{\polbook,\fquevedo}.

\vskip 1truecm
\noindent
{\it Continuous and Discrete Symmetries}

In building models coming from string theory, {\it there are no global internal
symmetries in spacetime} (there are no continuous global symmetries in all string theories). 
This is because if there is
an internal symmetry, there should be a vector field in the spectrum because the
properties of SCFT and it has the same properties of the gauge field of that
symmetry.

Take for instance Type I or Type II superstring theory. We know from Noether's theorem 
of the two-dimensional theory that
associated with the Type I or II  superconformal symmetry there is a worldsheet conserved supercharge,
$Q={1\over 2 \pi i}\int(dz d \theta J -d\bar{z}d\bar{\theta}\bar{J})$, where, by
uses of this symmetry, $J$ should be a $({1\over 2},0)$ tensor superfield and
$\bar{J}$ is a $(0,{1\over 2})$ tensor superfield. The associated bosonic vertex
operators
(when we combine these tensors with the fermionic fields $\widetilde{\psi}^{I}$ and $\psi^{I}$
respectively) have the property to couple with left and (or) right-moving parts of
$Q$, giving rise to a spacetime gauge symmetry. 

The absence of internal global symmetries in spacetime physics coming from string theory 
may help to understand some exciting problems of particle physics like the existence of the non-zero
neutrino masses, which lie in the violation of the leptonic number.  This was argued recently by
Witten in Ref. \neutrino.

However there are generically some discrete symmetries in string models. For
example, T-duality which is an infinite dimensional one, or some models inherited
from the point group of orbifold constructions which are finite dimensional, are in
fact regarded as discrete symmetries.
The importance of this kind of symmetries relay in the fact that they are useful for model
building,
hierarchy of masses and other related problems. 

\vskip 1truecm
\noindent
{\it P, C, T Symmetries}

We will see how discrete spacetime symmetries $P$, $C$ and $T$ are broken in string
theory. If string theory is correct, when we compactify (for example on  
Calabi-Yau manifolds, orbifolds, tori , etc.) one must obtain
the same symmetries (or broken symmetries) as these of the SM.

\item{$\bullet$}{\it P-symmetry}

Parity symmetry is violated by gauge interactions in SM. In string
theory there are an analogous situation. Take for instance the heterotic string. The
massless states in ten dimension are labeled by irreps of the little group SO(8). The action of
parity symmetry reverses the spinor representations ${\bf 8_s}$ and ${\bf 8_s}'^*$ of the
left and right-moving sectors. The   
symmetry is realized if the corresponding gauge representations ${\cal R}$ and ${\cal R}'^*$ 
are equal.

However, ${\cal R}$ is the adjoint representation while that  ${\cal R}'^*$ is empty. 
This tell us that parity symmetry is broken and the gauge couplings are chiral. But 
although in ten dimensions the spectrum is chiral, when we compactify to
four dimensions, the spectrum could be turned out into a non-chiral one.

For example, for toroidal compactifications, the spectrum is no-chiral, but for {\bf
$Z_3$} orbifold (compactification) the spectrum it is. Other kinds of   
compactifications produce chiral gauge couplings.

The chirality of the spectrum can be expressed by a topological quantity called {\it
Index} as we saw in the last subsection. Since the index is a topological invariant quantity,   
it does not suffers any change under continuous transformations of the CFT.

\item{$\bullet$}{\it C  and CP Symmetry} 

Charge conjugation symmetry is also broken in SM. This is
because $C$ leaves spacetime invariant, but conjugates the gauge generators. As in SM,
in string theories we require that conjugate representations (for example in SM,
the fermions) satisfy ${\cal R}_{\pm}={\cal R}^*_{\pm}$. From this we can see that chiral
gauge couplings do not satisfy $C$ symmetry. For the orbifold example this is also   
true.

Consider now the $CP$ symmetry. This symmetry takes ${\cal R}_+ \rightarrow {\cal R}^*_-$. Thus,
any gauge coupling satisfies it as a consequence of $CPT$ invariance. In the case of the orbifold 
there is a symmetry of
the action which reverses $X^{k}$ into $\psi^{k}$ with $k=3,5,7,9$, and all the
$\lambda^I$ ($I$ odd). This is a $CP$ symmetry in 4-dimensions. So, $Z_3$ orbifold is
$CP$ symmetric.

\item{$\bullet$}{\it CPT Symmetry} 

In string theory, as in local Lorentz-invariant quantum field
theory, $CPT$ symmetry is preserved.  In string perturbation theory we use the
$\theta$-operation\foot{This is basically, the same argument used in field
theory to prove $CPT$ symmetry.}. This is defined as $\theta(X^{0,3})=\psi^{0,3}$ and
vice versa. In Euclidean time this can be represented by a $\pi$-rotation in the
plane $(iX^0,X^3)$. In this context is clear that the action of $\theta$ is a
symmetry, that reverses time and includes parity (in $X^3$). In order to show that
this action also includes charge conjugation consider the $S$-matrix,

\eqn\cinco{
\langle \alpha , out|\beta ,in \rangle=  \langle \bar{\cal{V}}_{\alpha}{\cal V}_{\beta}\rangle ,
}
where ${\cal V}$ is the vertex operator and we are only considering vertex operators  
to the initial and final states. The action of $\theta$ is

\eqn\cseis{
\langle \alpha , out|\beta , in \rangle = \langle \theta \cdot \bar{{\cal V}}_{\alpha} \theta \cdot
{\cal V}_{\beta} \rangle= \langle \theta \bar{\beta}, out|\theta \bar{\alpha}, in \rangle.
} 
When we apply CPT operation, we can see that it is antiunitary and it is $\theta$
combined with the conjugation of the vertex operator

\eqn\csiete{
\langle CPT \cdot \beta, out|CPT \cdot \alpha , in \rangle = \langle \alpha , out|\beta , in 
\rangle.
}

The manner in what we saw that $CPT$ is an exact symmetry in string theory is only
applicable to the perturbative sector. However for the non-perturbative sector, we
can argue that SM, or field theory is the low energy limit of the string theory, so
we take $CPT$ symmetry for this low energy limit, and then we put it in 10 dimensions.

\vskip 1truecm
\noindent
{\it Effective Actions in Four Dimensions}

First of all, it is important to emphasize that consistent four-dimensional
superstring models which are chiral lead to ${\cal N}=1$ supersymmetric theories. 
At ${\cal N} \geq 2$ supersymmetry spoils chirality. Thus in order to consider
phenomenologically string models in four dimensions we restrict ourselves to construct ${\cal N}=1$
supersymmetric actions. Mostly of the material we describe here is at the review by F. Quevedo 
\fquevedo\ and we recommend it for checking details.

The corresponding spectrum of massless
particles are composed by graviton-gravitino multiplet ($G_{IJ}, \psi^{I})$
and
the gauge-gaugino multiplet $(A^{\alpha}_{I}, \lambda^{\alpha})$. Also there are
matter and moduli fields, in the form of chiral multiplets $(Z, \chi)$. In the
case of the dilaton field $\Phi$, it couples to the antisymmetric tensor $B_{IJ}$ to
form
the linear ${\cal N}=1$ multiplet $(\Phi , B_{IJ},\rho)$. We can construct a 
${\cal N}=1$ chiral
multiplet $(S, \chi_S)$ from the linear multiplet, where $S$ is obtained by a duality transformation of
the
dilaton field. This transformation is given by $S= a + ie^{\Phi}$ and $\nabla_{I}a
\equiv \varepsilon_{IJKL}\nabla^J B^{KL}$.

Thus the whole theory can be described as a ${\cal N}=1$ supergravity theory coupled
only to gauge and chiral multiplets. The most general Lagrangian which describes
these fields depends on three arbitrary functions of the chiral multiplets. These
fields are: $(i)$  $K(Z,Z^*)$, the K\"{a}hler potential which is a real function
which determines the kinetic terms of the chiral fields. The corresponding Lagrangian
is given by $L_{kin}=K_{ZZ^*}\partial_I Z \partial^I Z^*.$
$(ii)$ $W(Z)$, the superpotential, which is a holomorphic function of the chiral  
multiplets. $(iii)$ $f_{ab}(Z)$, the gauge kinetic function (holomorphic), and
determines the gauge kinetic couplings in the corresponding Lagrangian 
$L_{gauge}={\rm Re} f_{ab}(Z) F^a_{IJ}F^{IJ b}+ {\rm Im} f_{ab}(Z) F^a_{IJ}\widetilde{F}^{IJb}$. 
This function contributes to gaugino masses. 

There are another quantity called the scalar potential $V= V_F + V_D$,
where $V_F(Z,Z^*)=exp({K \over M^2_{Pl}}) \{ D_Z W
K^{-1}_{ZZ^*}{D_Z W}^*-3 {|W|^2 \over M^2_{Pl}} \}$,
$D_Z W = W_Z + W{K_Z \over M^2_{Pl}}$ and $V_D=({\rm Re} f^{-1})_{ab}(K_Z,
T^a Z)(K_{{Z}^*},T^b {Z}^*)$.

Thus, the problem to find an effective four-dimensional action, is to calculate the
functions $K,W$ and $f_{ab}$ when we are giving a specific string model. To do this, we have to
use all the symmetries we have and taking into account that four dimensional string
models are governed by two perturbation expansions. That is, an expansion in the 
sigma model (controlled by the size of the extra dimensions) and the proper string 
perturbation (in terms of the dilaton field) of the string coupling constant $g_S$.

First of all we consider only couplings generated
at string tree-level. For the sigma model, we also take only the tree-level
expansion. Using symmetries as the four dimensional Poincar\'e symmetry, supersymmetry,
gauge
symmetries and the axionic symmetry, we can extract the dependence of the effective
action on the dilaton field $S$. Then, at tree-level, the functions $K,W,f$ are
given by $K=-log(S+S^*)+\widehat{K}(T,U,Q)$, $W=Y_{IJK}Q^IQ^JQ^K$ and
$f_{ab}=S\delta_{ab}$ with $\widehat{K}$ an undetermined function. 
Our purpose is to find approximated expressions to this  
functions in the
tree-level of the string perturbation theory, but otherwise exact in the CFT.

It can be showed that by using the  axionic symmetries that at all orders in sigma-model
expansion, superpotential $W$ does not depend on $T$ and  $U$, so it is just a
function of the matter fields $Q^I$. Thus $W$ does not admit any kind of
corrections in the sigma model\foot{The reason for this, is that the field $T$,
related with the size of the extra dimensions, comes from the internal components
of the metric and determines the form that the loop expansion of the worldsheet
action takes.}.
The superpotential $W$ does not depends on $S$ as well. We know that $S$ is the string 
loop-counting parameter and this implies that $W$ is also an exact expression at
tree-level string perturbation theory, {\it i.e.,} does not admit any radiative
corrections.

Now, we are interested in finding a useful expression for $K$. This is more difficult,
because we only can calculate it for some simple cases. Take for example a
Calabi-Yau compactification whit $h_{1,1}=1$ and $h_{2,1}=0$. This give us that   
$K=-log(S+S^*)-3log(T+{T}^*+Q {Q}^*)$. When we write the K\"ahler potential as an
expansion in matter fields, it is possible to extract an exact tree-level
expression. The expansion is given by

$$
K=-log(S+ {S}^*)+K^M(T,{T}^*,U,{U}^*)+K^Q(T,{T}^*,U,{U}^*)Q{Q}^*
$$
$$
+\widehat{Z}(T,{T}^*,U,{U}^*)(QQ+Q^*Q^*)+{\cal O}(Q^3).
$$ 
For some  $(2,2)$ orbifold and Calabi-Yau models, it has been computed  the quantities
$K^M,K^Q$ and $\widehat{Z}$.

Consider now the loop corrections. We have seen that the superpotential (which is an
holomorphic function) does not admit radiative corrections. However, for the
K\"ahler potential this is different. We have just to calculate order by order in
the loop expansions the corresponding expression for $K_{loop}$. On the other hand,
the gauge kinetic function $f_{ab}$ is also holomorphic and we know the expression in an
exact manner for the tree level. Loop corrections to this function have a great 
importance due to this function determines the gauge coupling. Here we do not get
expressions for this corrections, but it is important to say that there are no
further corrections to $f_{ab}$ beyond one loop, as in the standard supersymmetric
theories.

In general, we have problems to determine how supersymmetry is broken at low
energies. We can not solve this problem within perturbative string theory. We need
work in the non-perturbative sector of the theory. But this sector, despite of many
efforts and excellent results, we do not yet have a complete
non-perturbative
version of string theory. However there are some interesting non-perturbative mechanisms 
to break supersymmetry as gaugino condensation, composite goldstinos and instantons.
The reader interested in these and another issues of non-perturbative string phenomenology
is encouraged to consult Refs. \refs{\ibanez,\dine}.

\vskip 2truecm
\newsec{T-duality, D-branes and Brane Configurations }

This section has the purpose of introducing basic ideas about T-duality in 
closed and open string theory. The open string case leads in a natural way to 
the definition of D-branes (for reviews of D-branes see Refs. \refs{\polbook,\pol,\cvj}). 
These objects are of extreme importance since
they are precisely the solitonic degrees of freedom which realize the strong-weak
coupling duality in superstring theory. This duality is also known as string S-duality.
T and S dualities relate the five perturbative superstring theories discussed previously 
and their
compactifications in diverse dimensions. Moreover, the
strong coupling limit of HE and Type II string theory (and their compactifications) suggest that
there is an
eleven-dimensional theory which has the eleven-dimensional supergravity as low energy
limit. This prospect of theory is widely known as M-theory. The name come from the words: mystery,
magic, mother, etc. Compactifications to diverse lower dimensions than ten gives more evidence 
of the existence of this theory. The fundamental degrees of freedom of this unified theory
are unknown, but macroscopically they include membranes and fivebranes. `Matrix Theory'
is an attempt to give the dof's of M-theory. The proposal is that these degrees of freedom
are the D0-branes. The worldvolume effective theory of a gas of $N$ D0-branes is a
SU$(N)$ quantum mechanics. Large $N$-limit reproduces the description of membranes and
fivebranes and some other results of eleven dimensional supergravity (for some reviews 
the reader can consult Refs. \refs{\banks,\taylor}). 

D-branes also, are very important tools to study the strong coupling of supersymmetric
theories in various dimensions. Different properties as chirality, dualities etc. are
encoded in the engineering of brane configurations. The moduli space of these susy
gauge theories is described by the Higgs and the Coulomb branches of the corresponding
brane configuration. Many field theory results are understanding in terms of a 
geometrical language and many generalizations have been established motivated by the 
brane engineering (more about this topic can be found in Ref. \gk). 

Finally, the presence of branes leads to modify the prescription of Calabi-Yau or orbifold
compactifications and new non-perturbative are possible. In these sections we 
will discuss some of these interesting topics.

\vskip 1truecm
\subsec{Toroidal Compactification, $T$-duality and D-branes}

D-branes are, despite of the dual fundamental degrees of freedom in string theory,
extremely interesting and useful tools to study nonperturbative  properties of
string and field theories (for some  reviews see \refs{\pol,\cvj}). Non-perturbative
properties of supersymmetric gauge theories can be better understanding as the
world-volume effective theory of some configurations of intersecting D-branes
(for a review see \gk). D-branes also are very important to connect gauge
theories with gravity. This is the starting point of the AdS/CFT correspondence
or Maldacena's conjecture. We don't review this interesting subject in this paper,
however the reader can consult the excellent review \malda.
Roughly speaking D-branes are static solutions of string equations which satisfy
Dirichlet boundary conditions. That means that open strings can end on them.
To explain these objects we follow the traditional way, by using T-duality on
open strings we will see that Neumann conditions are turned out into the Dirichlet ones.
To motivate the subject we first consider T-duality in closed bosonic string theory.

\vskip 1truecm
\noindent
{\it T-duality in Closed Strings}

The general solution of Eq. (3.5) in the conformal gauge can be written
as $X^I (\sigma , \tau ) = X^I_R(\sigma^-)+X^I_L (\sigma^+)$, where
$\sigma^{\pm}=\sigma \pm \tau$. Now, take one coordinate, say $X^{25}$ and compactify it on a
circle of radius $R$. Thus
we have that $X^{25}$ can be identified with $X^{25} + 2\pi R m$ where $m$ is called the {\it
winding
number}. The general solution for $X^{25}$ with the above compactification condition is

$$
X^{25}_R(\sigma^-)=X^{25}_{0R} +\sqrt{{\alpha ' \over 2}}P^{25}_R(\tau -\sigma 
)+i\sqrt{{\alpha ' \over 2}}\sum_{l \neq 0}{1\over l}\alpha^{25}_{R,l}exp
\bigg(-il(\tau - \sigma )\bigg)$$

\eqn\cocho{
X^{25}_L(\sigma^+)=X^{25}_{0L}+\sqrt{{\alpha '\over 2}}P^{25}_L(\tau +
\sigma)+i\sqrt{{\alpha '\over 2}}\sum_{n \neq
0}{1 \over l}\alpha^{25}_{L,l}exp\bigg(-il(\tau +\sigma )\bigg),}
where
\eqn\cochouno{
P^{25}_{R, L}={1 \over \sqrt{2}} \bigg({\sqrt{\alpha '} \over R}n \mp
{R\over \sqrt{\alpha '}}m\bigg).}

Here $n$ and $m$ are integers representing the discrete momentum and the winding
number, respectively. The latter has not analogous in field theory.  
While the canonical momentum is given by $P^{25}={1\over \sqrt{2\alpha
'}}(P^{25}_L+P^{25}_R)$. Now, by the mass shell condition, the mass of the perturbative states
is given by
$M^2=M_L^2+M_R^2$, with

\eqn\cnueve{
M^2_{L,R}=-{1 \over 2}P^I P_I ={1\over 2}(P^{25}_{L,R})^2+{2\over \alpha '}(N_{L,R}-1).}

We can see that for all states with $m \neq 0$, as $R \rightarrow \infty$ the mass
become infinity, while  $m = 0$ implies that the states take all values for $n$ and form a
continuum. At the case when $R \rightarrow 0$, for states with $n \neq 0$, mass
become infinity. However in the limit $R \rightarrow 0$ for $n =0$ states with all
$m$ values produce a continuum in the spectrum. So, in this limit the compactified
dimension disappears. For this reason, we can say that the mass spectrum of the theories at radius
$R$
and ${\alpha '\over R}$ are identical when we interchange $n \Leftrightarrow m$. This
symmetry is known as  {\it T-duality}.

The importance of T-duality lies in the fact that the T-duality transformation is a parity
transformation acting on the left and right moving
degrees of freedom. It leaves invariant the left movers and changes the sign of the right movers
(see Eq. (4.2))
\eqn\cincuenta{
P^{25}_L \rightarrow P^{25}_L, \ \ \ \ \ \ \ \
P^{25}_R \rightarrow -P^{25}_R.}

The action of T-duality transformation must leave invariant the whole theory
(at all order in perturbation theory). Thus, all kind of interacting states in certain theory
should correspond to those states belonging to the dual theory. In this context,   
also the vertex operators are invariant. For instance the tachyonic vertex operators are

\eqn\ciuno{
{\cal V}_L= exp(iP_L^{25}X^{25}_L), \ \ \ \ \ \ \ \ \
{\cal V}_R= exp(iP^{25}_RX^{25}_R).}
Under T-duality, $X^{25}_L \rightarrow X^{25}_L$ and $X^{25}_R \rightarrow
-X^{25}_R$; and from the general solution Eq. (13), ${\alpha}^{25}_{R,i} \rightarrow  
-{\alpha}^{25}_{R,i}$, $X^{25}_{0R} \rightarrow -X^{25}_{0R}$. Thus, T-duality
interchanges $n \Leftrightarrow m$ (Kaluza-Klein modes $\Leftrightarrow$ winding
number) and $R \Leftrightarrow {\alpha ' \over R}$ in closed string theory.

\vskip 1truecm

\noindent
{\it T-duality in Open Strings}

Now, consider {\it open strings} with Neumann boundary conditions. Take again the
$25^{th}$ coordinate and compactify it on a circle of radius $R$, but keeping Neumann
conditions. As in the case of closed string, center of mass momentum takes only discrete
values $P^{25}={n\over R}$. While there is not analogous for the winding
number. So, when $R \rightarrow 0$ all states with nonzero momentum go to infinity 
mass, and do not form a continuum. This behavior is similar as in field theory, but
now there is something new. The general solutions are

$$
X^{25}_R ={X_0^{25} \over 2} -{a\over 2} + \alpha 'P^{25}(\tau -\sigma
) +i\sqrt{{\alpha '\over 2}}\sum_{l \neq 0}{1\over l} \alpha^{25}_l exp \bigg(-i2l(\tau
-\sigma ) \bigg),$$

\eqn\cidos{
X^{25}_L ={X_0^{25}\over 2} +{a\over 2} +\alpha 'P^{25}(\tau +\sigma
)+i\sqrt{{\alpha ' \over 2}}\sum_{l \neq 0}{1 \over l}\alpha^{25}_l exp \bigg(-i2l(\tau
+\sigma ) \bigg)}
where $a$ is a constant.
Thus,
$X^{25}(\sigma ,\tau ) = X^{25}_R(\sigma^-) +X^{25}_L(\sigma^+)=X_0^{25}
+{2\alpha 'n \over R}\tau +  oscillator \ terms.
$
Taking the limit $R \rightarrow 0$, only the $n=0$ mode survives. Because of this,
the string seems to move in 25 spacetime dimensions. In other words, the strings vibrate in 24
transversal directions. T-duality provides a new T-dual coordinate defined by
$\widetilde{X}^{25}(\sigma ,\tau
)=X^{25}_L(\sigma^+)-X^{25}_R(\sigma^-)$. Now, taking
$\widetilde{R}={\alpha
' \over R}$ we have $\widetilde{X}^{25}(\sigma ,\tau )=a +2 \widetilde{R} \sigma n + oscillator \
terms.$
Using the boundary conditions at $\sigma =0,\pi$ one has
$ \widetilde{X}^{25}(\sigma ,\tau ) \mid_{\sigma =0} = a$ and $
\widetilde{X}^{25}(\sigma ,\tau )\mid_{\sigma =\pi}=a +2\pi \widetilde{R}n.$   
Thus, we started with an open bosonic string theory with Neumann boundary
conditions, and T-duality and a compactification on a circle in the $25^{th}$
dimension, give us Dirichlet boundary conditions in such a coordinate. We can
visualize this saying that an open string has its endpoints fixed at a hyperplane
with 24 dimensions.

Strings with $n=0$ lie on a 24 dimensional plane space (D24-brane). Strings with
$n=1$ has one endpoint at a hyperplane and the other at a different hyperplane which
is separated from the first one by a factor equal to $2\pi \widetilde{R}$, and so on. But
if we compactify $p$ of the $X^i$ directions over a $T^{p}$ torus ($i=1,...,p$). Thus, after
T-dualizing
them we have strings with endpoints fixed at hyperplane with $25-p$ dimensions, the D$(25-p)$-brane.

Summarizing: the system of open strings moving freely in spacetime with $p$ compactified
dimensions on $T^p$ is equivalent, under T-duality, to strings whose enpoints are fixed at a
D$(25-p)$-brane {\it i.e.} obeying Neumann boundary conditions in the $X^i$ longitudinal
directions ($i=1,\dots ,p$) and Dirichlet ones in the transverse coordinates $X^m$
($m=p+1,...,25$).

The effect of T-dualizing a coordinate is to change the nature of the boundary conditions, from
Neumann  to Dirichlet and vice versa. If one dualize a longitudinal coordinate this coordinate
will
satisfies the Dirichlet condition and a D$p$-brane becomes a D$(p-1)$-brane. But if the
dualized coordinate is one of the transverse coordinates the D$p$-brane becomes a
D$(p+1)$-brane.

T-duality also acts conversely. We can think to begin with
a closed string theory, and compactify it on to a circle in
the $25^{th}$ coordinate, and then by imposing Dirichlet conditions, obtain a
D-brane. This is precisely what occurs in Type II theory, a theory of closed strings.

\vskip 1truecm

\noindent
{\it Spectrum and Wilson Lines}

Now, we will see how does emerges a gauge field on the D$p$-brane world-volume. Again, for  
the mass
shell condition for open bosonic strings and because T-duality
$M^2=({n \over \alpha '}\widetilde{R})^2+{1\over \alpha '}(N-1)$. The massless state
($N=1$, $n=0$) implies that the gauge boson ${\alpha}^{I}_{-1}\mid 0 \rangle$ ($U(1)$
gauge boson) lies on the D24-brane world-volume. On the other hand,
${\alpha}^{25}_{-1}\mid 0 \rangle$ has a {\it vev} (vacuum expectation value) which describes
the position $\widetilde{X}^{25}$ of the D-brane after T-dualizing. Thus, we can say
in general, there is a gauge theory $U(1)$ over the world volume of the D$p$-brane.

Consider now an {\it orientable open string}. The endpoints of the string carry
charge under a non-Abelian gauge group. For Type II theories the gauge group is
$U(N)$. One endpoint transforms under the fundamental
representation ${\bf N}$ of $U(N)$ and the other one, under its complex conjugate representation
(the anti-fundamental one) ${\bf N}^*$.

The ground state wave function is specified by the center of mass momentum and by
the charges of the endpoints. Thus implies the existence of a basis $\mid k; ij \rangle$
called {\it Chan-Paton basis}. States $\mid k;ij \rangle$ of the Chan-Paton basis are those
states which carry charge 1 under the $i^{th}$ $U(1)$ generator and $-1$ under the
$j^{th}$ $U(1)$ generator. So, we can decompose the wave function for ground state
as $\mid k;a \rangle=\sum_{i,j=1}^N \mid k;ij \rangle \lambda^a_{ij}$ where $\lambda^a_{ij}$ are
called {\it Chan-Paton factors}. From this, we see that it is possible to add degrees of freedom
to endpoints of the string, that are precisely the Chan-Paton factors.

This is consistent with the theory, because the Chan-Paton factors have a Hamiltonian
which do not posses dynamical structure. So, if one endpoint to the string is
prepared in a certain state, it always will remains the same. It can be deduced from
this, that $\lambda^a \longrightarrow U\lambda^aU^{-1}$ with $U\in$ $U(N)$. Thus, the worldsheet
theory is
symmetric under $U(N)$, and this global symmetry is a gauge symmetry in spacetime. So the
vector state at massless level ${\alpha}^{I}_{-1}\mid k,a \rangle$ is a $U(N)$ gauge boson.

When we have a gauge configuration with non trivial line integral around a
compactified dimension (i.e a circle), we said there is a Wilson line. In case of
open strings with gauge group $U(N)$, a toroidal compactification of the $25^{th}$
dimension on a circle of radius $R$. If we choice a background field $A^{25}$ given
by $A^{25} ={1\over 2\pi R} diag(\theta_1,...,\theta_N)$ a Wilson line
appears. Moreover, if $\theta_i=0$, $i=1,...,l$ and $\theta_j \neq 0$, $j=l+1,...,N$
then gauge group is broken: $U(N) \longrightarrow U(l) \times U(1)^{N-l}$. It is 
possible to deduce that $\theta_i$ plays the role of a Higgs field.
Because string states with Chan-Paton quantum numbers $\mid ij \rangle$ have charges $1$   
under $i^{th}$ $U(1)$ factor (and $-1$ under $j^{th}$ $U(1)$ factor) and neutral with
all others; canonical momentum is given now by $
P^{25}_{(ij)} \Longrightarrow {n\over R} + {(\theta_j -\theta_i ) \over 2\pi R}.
$
Returning to the mass shell condition it results,

\eqn\citres{
M^2_{ij}=\bigg({n\over R} +{\theta_j - \theta_i \over 2\pi R}\bigg)^2 +{1\over \alpha '}(N-1).}

Massless states ($N=1, n=0$) are those in where $i=j$ (diagonal terms) or for which
$\theta_j=\theta_i$ $(i \neq j)$. Now, T-dualizing we have
$
\widetilde{X}^{25}_{ij}(\sigma ,\tau )=a +(2n
+{\theta_j- \theta_i \over \pi})\widetilde{R}\sigma + oscillator \ terms.$ 
Taking $a=\theta_i \widetilde{R}$,
$
\widetilde{X}^{25}_{ij}(0,\tau )= \theta_i \widetilde{R}
$ and
$
\widetilde{X}^{25}_{ij}(\pi ,\tau )= 2\pi n \widetilde{R} + \theta_j \widetilde{R}.
$
This give us a set of $N$ D-branes whose positions are given by
$\theta_j\widetilde{R}$, and each set is separated from its initial positions
($\theta_j=0$) by a factor equal to $2\pi \widetilde{R}$.
Open strings with both endpoints on the same D-brane gives massless gauge bosons.
The set of $N$ D-branes give us $U(1)^N$ gauge group. An open string with one endpoint
in one D-brane, and the other endpoint in a different D-brane, yields a
massive state with $M \sim (\theta_j -\theta_i)\widetilde{R}$. Mass decreases when two
different D-branes approximate to each other, and are null when become the same.
When all D-branes take up the same position, the gauge group is enhanced from $U(1)^N$ to
$U(N).$ On the D-brane world-volume there are also scalar fields in the adjoint representation
of the gauge group $U(N)$. The scalars parametrize the transverse positions of the D-brane in the
target space $X$.

\vskip 1truecm

\noindent
{\it D-Brane Action}

With the massless spectrum on the D-brane world-volume it is possible to construct a low
energy effective action. Open strings massless fields are interacting with the closed
strings massless spectrum from the {\bf NS-NS} sector. Let $\xi^a$ (with $a=0, \dots ,
p$) be the world-volume coordinates on $W$. The effective action is the gauge invariant
action well known as the Dirac-Born-Infeld (DBI)-action

\eqn\cicuatro{
S_D = - T_p \int_W d^{p+1} \xi e^{-\Phi}        \sqrt{ det\big( G_{ab} + B_{ab} + 2 \pi
\alpha ' F_{ab} \big)},}
where $T_p$ is the tension of the D-brane, $G_{ab}$ is the world-volume induced metric,
$B_{ab}$ is the induced antisymmetric field, $F_{ab}$ is the Abelian field strength on
$W$ and $\Phi$ is the dilaton field.

For $N$ D-branes the massless fields turns out to be $N \times N$ matrices and the
action turns out to be non-Abelian DBI-action (for a nice review about the Born-Infeld
action in string theory see \tsey) 

\eqn\cicinco{
S_D = - T_p \int_W d^{p+1} \xi e^{- \Phi} Tr\bigg(\sqrt{ det\big( G_{ab} + B_{ab}   
+ 2 \pi \alpha ' F_{ab}\big)} + O\big( [X^m,X^n]^2 \big) \bigg)}
where $m,n = p+1, \dots , 9$. The scalar fields $X^m$ representing the transverse
positions become $N \times N$ matrices and so, the spacetime become a noncommutative
spacetime. We will come back later to this interesting point.

\vskip 1truecm

\noindent
{\it Ramond-Ramond Charges}

D-branes are coupled to Ramond-Ramond (RR) fields $G_p$ \pol. The complete
effective action on the D-brane world-volume $W$ which taking into account this coupling
is

\eqn\ciseis{
S_D = - T_p \int_W d^{p+1} \xi \bigg\{e^{-\Phi} \sqrt{ det\big( G_{ab} + B_{ab} + 2 \pi
\alpha ' F_{ab} \big)} + i \mu_p \int_W \sum_p C_{(p+1)} Tr\bigg( e^{2 \pi \alpha '
(F+B)}\bigg) \bigg\}}
where $\mu_p$ is the RR charge. RR charges can be computed by considering the anomalous
behavior of the action at intersections of D-branes. Thus RR charge is given
by

\eqn\cisiete{
Q_{RR} = ch(j!E) \sqrt{ \widehat{A}(TX)},}
where $j : W \hookrightarrow X$. Here $E$ is the Chan-Paton bundle over $X$,
$\widehat{A}(TX)$ is the genus of the spacetime manifold $X$. This gives an ample evidence
that the RR charges take values not in a cohomology theory, but in fact, in a K-Theory. 
This result was developed by Witten in Ref. \k\ in the context of non-BPS brane
configurations worked out by Sen \asen.  

Finally, RR charges and RR fields do admit a classification in terms of topological 
K-theory. The inclusion of a $B$-field turns out the effective theory non-commutative and
a suitable generalization of the topological K-theory is needed. The {\it right}
generalization seems to be the K-Homology and the K-theory of $C^*$ algebras
\kk. This subject is 
right now under intensive investigation.

\vskip 1truecm
\subsec{D-brane Configurations and Susy Gauge Theories}

As an application of the D-brane theory we consider in this section
the brane box models, in particular we focus on the {\it cube} model discussed in 
\cube.

The dynamics of D-branes in certain configurations of intersecting branes
encodes many field-theoretical facts about supersymmetric theories in   
several dimensions (for a review, see \gk). Gauge theories in $(p+1)$
dimensions with
sixteen supercharges can be obtained as the world-volume theories of flat
infinite Dp-branes. In the context of theories with eight supersymmetries
in $p$ dimensions, it was shown in \hw\ that such theories can be realized
by considering  Dp-branes with a world-volume which is finite in one
direction, in which the D-brane ends on NS fivebranes. 
The brane is suspended between NS 
fivebranes spanning 012345. The low
energy theory in the non-compact dimensions of the D-brane is
$p$-dimensional. It is still a gauge theory, but the presence of the NS
branes breaks half of the supersymmetries, so eight supercharges remain.
This construction has been generalized in several directions, and has
yielded
the realization of a large family of models in several dimensions. This
setup has also been exploited to compute different exact quantum results   
in these theories. For a review of such achievements, see \gk.

A nice property of the interplay of field theories and configurations of
branes is that the intersections of branes
can sometimes support chiral zero modes. This opens the possibility of   
studying chiral gauge theories using branes. The simplest such example is
provided by the realization of six-dimensional theories with eight
supersymmetries, which are chiral. These can be realized in the
setup described above by taking $p=6$, {\it i.e.} one considers D6 branes
extending along 0123456, and which are bounded in 6 by NS branes with
world-volume along 012345.

Chirality if a fragile property, in the sense that toroidal
compactifications or too much supersymmetry spoil it. Thus, in order to 
obtain chiral theories in four dimensions one has to consider theories with
only four supercharges. Their realization in terms
of branes requires new ingredients. A fairly general family of brane
configurations realizing generically chiral gauge theories in four
dimensions was introduced  in Ref. \hz.

The idea is a clever extension of the 
philosophy in \hw. It consists in realizing first a five-dimensional 
theory with eight supercharges, by using D5 branes along 012346,
suspended between NS branes with world-volume along 012345. Then, the D5
brane is bounded
in the direction 4, by using a new set of NS branes oriented along 012367
(denoted NS$'$ branes). The low energy theory is four-dimensional, since
the world-volume of the D5 brane  along 46 is a finite rectangle. Such
configurations are known as brane box models. The presence
of the new kind of branes breaks a further half of the supersymmetries,
and so the theory has only four supercharges. Furthermore, the
intersections of NS, NS$'$ and D5 branes introduce chirality in the
four dimensional theory. There is no complete understanding of the
quantum effects of these gauge theories in terms of branes.

\vskip 1truecm
\noindent
{\it The Cube Brane Box Configurations and Susy Theories in Two Dimensions}

Here we introduce certain supersymmetric configurations of
NS, NS$'$, and NS$''$ branes, and D4 branes in Type IIA superstring
theory. They give rise to two-dimensional $(0,2)$
field theories. These configurations are obtained
in the spirit of the brane box configurations in \hz, by considering
D-branes which are finite in several directions. As explained before, they 
belong to a natural sequence of brane box models
yielding chiral theories in six, four and two dimensions (taking D branes
compact in one, two and three directions, respectively).

Let us consider the ingredients of the brane configurations which we will
use in this paper. Brane configurations consist of:

\item{$\bullet$} NS fivebranes located  along $(012345)$.

\item{$\bullet$} NS' fivebranes located  along $(012367)$.

\item{$\bullet$} NS'' fivebranes located  along $(014567)$.

\item{$\bullet$} D4 branes located along $(01246)$.

In this configuration the D4 branes are finite in the directions
246. They are bounded in the direction $2$ by the NS$''$ branes, in
the direction $4$ by the NS$'$ branes, and in the direction $6$ by the NS
branes. For the D4 branes to be suspended in this way, it is necessary
that the coordinates of all branes in $89$ should be equal. It is also
required that  two NS branes joined by a D4 brane should have the
same position in 7, and analogously that two NS$'$ branes joined by a
D4 brane should have the same
position in 5, and that two NS$''$ branes should have the same
position in 3. 

The low-energy effective field theory on the D4 branes is two-dimensional,
since $01$ are the only non-compact directions in their world-volume. The
presence of each kind
of NS fivebrane breaks one half of the supersymmetries, and altogether
they break to $1/8$ of the original supersymmetry. A further half is
broken by the D4 branes, and the world-volume theory has
$(0,2)$ supersymmetry in two dimensions. Since the D brane is bounded by
NS fivebranes, the world-volume gauge bosons will not be projected out and
there will be a gauge group for each box in the model.
The $U(1)_R$ R-symmetry of the field theory is manifest as the rotational
symmetry in the directions 89.

We note that there are a variety of other objects that can be introduced
in the configuration without breaking the supersymmetry. For instance,
there are three kinds of D6 branes that can be introduced, namely D6
branes along 0124789, D6$'$ branes along 0125689, and D6$''$ branes along
0134689. They provide vector-like flavors for the gauge groups. These
extensions are quite well-known from other contexts, and we will not study
them in the present paper.

There is a first rough classification we can make in these brane
configurations, according to whether the directions 246 are taken compact 
or not. If some of these directions are non-compact, then there will be
some semi-infinite box, which will represent some global symmetry. For
definiteness we will center on the case in which all three directions are
compact, with lengths $R_2$, $R_4$ and $R_6$. Extension of our results to
other cases is straightforward.

A generic configuration consists of a three-dimensional grid of $k$ NS
branes, $k'$ NS' branes and $k''$ NS'' branes dividing the 246
torus into a set of $kk'k''$ boxes. We will often think about these
configurations as infinite periodic arrays of boxes in $\IR^3$,
quotiented by an infinite discrete group of translations in a
three-dimensional lattice $\Lambda$. This point
of view is particularly useful to define models in which the unit cell has
non-trivial identifications of sides \hu. 

\vskip 1truecm
\noindent
{\it (0,2) Effective Theory on the D4 Brane}

The effective field theory on the only non-compact directions $01$ of the D4 branes 
world-volume is a (0,2) gauge theory
in two dimensions. These theories are described in the (0,2)
superspace $(y^{\alpha},\theta^+,\bar{\theta}^+)$. There are three basic
kinds of multiplets which we will use.

\item{$\bullet$} The $(0,2)$ {\it gauge multiplet} $V'$, which contains gauge bosons
$v_{\alpha}$, $\alpha=0,1$, and one fermion $\chi_-$.

\item{$\bullet$} The $(0,2)$ {\it chiral multiplet} $\Phi'$, contains one complex
scalar $\phi$ and one chiral fermion $\psi_+$.

\item{$\bullet$} The $(0,2)$ {\it Fermi multiplet}, $\Lambda$, is described
by an anticommuting superfield. Its complete $\theta$ expansion contains a
chiral spinor $\lambda_-$,
an auxiliary field $G$, and a holomorphic function $E$ depending on the
chiral (0,2) superfields $\Phi_i'$. The Fermi multiplet  $\Lambda$
satisfies the constraint $\bar{\cal D}_+ \Lambda= \sqrt{2} E(\Phi')$,
with $\bar{\cal D}_+ E = 0$. Here $\bar{\cal D}_+$ represents the supersymmetric
covariant derivative. The expansion in components for the Fermi superfield
is

\eqn\ciocho{
\Lambda=\lambda_- -\sqrt{2}\theta^+ G - i\theta^+ {\bar
\theta}^+ (D_0+D_1)\lambda_- - \sqrt{2} {\bar \theta}^+ E(\Phi')}
with $D_{\alpha}$ denoting the usual supersymmetric derivative.

Gauge theories involving these fields are described by a Lagrangian with
the 
\eqn\cinueve{ 
L =  L_{gauge} + L_{ch} + L_F + L_{D,\theta} + L_J .}
As usual, $L_{gauge}$ is the kinetic term of the gauge multiplet given by   

\eqn\sesenta{
L_{gauge} = {1 \over 8g^2} \int d^2y d\theta^+ d \bar{\theta}^+
{\rm Tr} \big(\bar{\Upsilon} \Upsilon\big)}
where $\Upsilon$ is the field strength of $V'$.

The term $L_{ch}$ contains the
kinetic energy and gauge couplings of the (0,2) chiral superfields
$\Phi'_i$. It is given by

\eqn\suno{
L_{ch} = - {i \over 2} \int d^2y d^2 \theta \sum_i
\bigg(\bar{\Phi'_i}({\cal D}_0 - {\cal D}_1) \Phi'_i \bigg),}
where ${\cal D}_{0}$ and ${\cal D}_{1}$ are the (0,2) gauge covariant derivatives
with respect to $V'$.

The term $L_F$ describes the dynamics of the Fermi multiplets $\Lambda$,
and certain interactions. It is given by

\eqn\sdos{
L_F= -{1\over 2} \int d^2yd^2\theta \sum_a \big(
\bar{\Lambda}_{a} {\Lambda}_{a} \big).}

Substitution of (4.12) into (4.16) leads to

\eqn\stres{
L_F\, =\, \int d^2y \sum_a \bigg\{
i{\bar\lambda_{-,a}}(D_0+D_1) 
\lambda_{-,a} + |G_a|^2 - |E_a|^2 - \sum_j\big( {\bar \lambda_{-,a}}
{\partial E_a
\over \partial \phi_j}\, \psi_{+,j} + {\partial \bar{E_a} \over \partial
\bar{\phi_j}}\, {\bar \psi_{+,j}} \lambda_{-,a} \big) \bigg\}.}

The Fayet-Iliopoulos and theta angle terms are encoded in the (0,2)
Lagrangian $L_{D,\theta}$ which is written as

\eqn\scuatro{
L_{D,\theta} = {t \over 4} \int d^2 y d\theta^+ {\rm Tr}
\big(\Upsilon
|_{\bar{\theta}^+ = 0} \big) + h.c.} 
   where $t = {\theta\over 2 \pi} + ir$.

Finally  (0,2)  models do admit an additional
interaction term $L_J$ which depends on a set of holomorphic functions
$J^a(\Phi')$ of the chiral superfields. There is one such function for
each Fermi
superfield. They satisfy the relation $\sum_a E_a J^a =0$.
This interaction is the (0,2) analog of the superpotential, and its
Lagrangian $L_J$ is given by  

\eqn\scinco{
L_J = -{1 \over \sqrt{2}} \int d^2 y d\theta^+
 \sum_a \bigg(\Lambda_{a}J^a|_{\bar{\theta}^+ = 0} \bigg) - h.c. \ \ .}

The expansion of this term in components is

\eqn\sseis{
L_J=-\int d^2y \sum_a \bigg( G_a J^a + \sum_j \lambda_{-,a}
\psi_{+,j}
{{\partial J^a} \over {\partial \phi_j}} \bigg) - h.c. \ \ .}

After combining the Lagrangians $L_F$ and $L_J$ and
solving for the equations of motion for the auxiliary fields $G$,
the relevant interaction terms in the Lagrangian (we are not listing  
the gauge interactions and D-terms here) are

\eqn\ssiete{
\sum_a \big( |J^a(\phi)|^2 + |E^a(\phi)|^2 \big)
- \sum_{a,j} \big( {\bar \lambda_{-,a}} {\partial E_a \over \partial
\phi_j} \psi_{+,j} + \lambda_a {\partial J^a \over \partial \phi_j}
\psi_{+,j} + h.c.\big).}

The first term contains the scalar potential, and the second the Yukawa 
couplings. Notice that the choice of the functions $E$ and $J$ completely
defines the interactions of the theory.

For (0,2) theories in two dimensions we have just one U(1) R-symmetry 
group acting on the superspace coordinates $(\theta^+,\bar{\theta}^+)$.
This is right-moving  R-symmetry and it acts as
$\theta^+ \to e^{i \beta} \theta^+$,
$\bar{\theta}^+ \to e^{-i \beta} \bar{\theta}^+$, leaving $\theta^-$ ,
$\bar{\theta}^-$ invariant.

\vskip 1truecm
\noindent
{\it Interpretation of the Linear Sigma Model}

Up to here we have introduced a large family of
two-dimensional $(0,2)$ gauge theories. Since $(2,2)$ and $(0,2)$ theories
have been traditionally used as world-sheet descriptions of string
theories propagating on some target space, it is a natural
question
whether the (classical) Higgs branch or our models has any geometrical
interpretation
of the kind. In this section we are to show that it describes the
dynamics of a type IIB D1 brane on a $IC^4/\Gamma$ singularity, with 
$\Gamma$ an abelian subgroup of $SU(4)$. The main tool for reaching this
conclusion will be a T-duality performed on the brane box model along the
directions 246.

In this section we perform a T-duality on the brane box models along the
directions 246. The main tool will be the well known T-duality relation
between a set of $n$ parallel NS fivebranes and $n$ Kaluza-Klein
monopoles. The discussion in this subsection parallels that in \hu.

Let us start with the simplest case of a brane box model formed by a unit
cell of $k \times k' \times k''$ boxes, with trivial identifications of
faces. In this case the T-duality along the directions 246 is
particularly  
easy. We start with $k$ NS branes along 012345, $k'$ NS$'$ branes along
012367, and $k''$ NS$''$ branes along 014567. The T-duality along 2
transforms the NS$''$ branes into $k''$ Kaluza-Klein monopoles. These will
be described by a multi-center Taub-NUT metric, with non-trivial geometry
on the directions 2$'$,3,8,9, with 2$'$ denoting the coordinate dual to 2.
Notice that, since the 3,8,9 coordinates of the initial NS$''$ branes
coincided, so do the coordinates of the corresponding $k''$ centers of the
Taub-NUT metric, so that it contains singularities of type
$A_{k''-1}$.

Similarly, the T-duality along 4 transforms the $k'$ NS$'$ branes into  
$k'$
Kaluza-Klein monopoles with world-volume along 012367, and represented by
a nontrivial geometry on 4$'$,5,8,9. Again, since the centers of the  
Kaluza-Klein monopoles coincide, such geometry will contain singularities
of type $A_{k'-1}$. Finally, the T-duality along 6 turns the $k$ NS
branes into $k$ Kaluza-Klein monopoles. Their world-volume spans 012345,
and they are represented by a non-trivial geometry along 6$'$,7,8,9. Since
again all the centers coincide, there will be $A_{k-1}$ singularities.

Thus, the final T-dual of the grid of NS, NS$'$ and NS$''$ branes is
type IIB string theory with a complicated geometry in the directions
2$'$,3,4$'$,5,6$'$,7,8,9. One can
think of it roughly as some `superposition' of the Kaluza-Klein monopoles 
we have described. Even without a quantitative knowledge of such metric,
we can describe the relevant features for our purposes. One such
feature is that the number of unbroken supersymmetries constrains
the manifold to be a Calabi-Yau four-fold. Also, from our remarks
above we know the existence of certain (complex) surfaces of
singularities of type $A_{k-1}$, $A_{k'-1}$ and $A_{k''-1}$ singularities.
If we introduce complex coordinates $w_1=\exp (x^7+ix^{6'})$
,$w_2=\exp(x^5+ix^{4'})$, $w_3=\exp(x^3+ix^{2'})$, and $w_4=x^9+ix^8$, the
surface of $A_{k-1}$ singularities is defined roughly by $w_1=w_4=0$, 
the surface of $A_{k'-1}$ singularities is defined by $w_2=w_4=0$, and the
surface of $A_{k''-1}$ singularities is given by $w_3=w_4=0$. At the
origin $w_1$$=w_2$$=w_3$$=w_4=0$ all surfaces meet and the singularity is
worse. It can be described as a quotient singularity of type
$\IC^4/\Gamma$, with $\Gamma=\IZ_k\times \IZ_{k'}\times \IZ_{k''}$.   
This
discrete group is generated by elements $\theta$, $\omega$, $\eta$, whose
action on $(z_1,z_2,z_3,z_4)\in \IC^4$ is as follows:

$$
\matrix{
\theta: & (z_1,z_2,z_3,z_4) & \to  & (e^{2\pi i/k}z_1,z_2,z_3,e^{-2\pi i/k}z_4) \cr
\omega: & (z_1,z_2,z_3,z_4) & \to  & (z_1,e^{2\pi i/k'}z_2,z_3,e^{-2\pi i/k'}z_4) \cr
\eta: & (z_1,z_2,z_3,z_4) & \to & (z_1,z_2,e^{2\pi i/k''}z_3,e^{-2\pi i/k''}z_4).  \cr
}
$$
In this description it becomes clear that there may be further surfaces of
singularities when the greatest common divisor of any two of $k,k',k''$ is
not $1$, in analogy with the discussion in \hu. This will not be
relevant for our purposes and we do not develop the issue further.   

After the T-duality, the initial D4 branes become D1 branes located at a
point in the four-fold. When the initial D4 branes are bounded by the grid
of NS, NS$'$ and NS$''$ branes, the T-dual  D1 branes will be located
precisely at the $\IC^4/\Gamma$ singular point. The field theories  
introduced previously correspond to the field theories
appearing in the world-volume of such D1 brane probes. In addition,
the structure of the singularity controls the spectrum and dynamics of the field theory
(for a recent review see \uranga.

Brane boxes models generating gauge theories in two dimensions with enchanced chiral (0,4),
(0,6) and (0,8) supersymmetry can be constructed and are also described at
\cube.

\vskip 2truecm

\newsec{Non-perturbative String Theory}

\subsec{Strong-Weak Coupling String Duality}

We have described the massless spectrum of the five consistent superstring
theories in ten dimensions. Additional theories can be constructed in lower dimensions
by compactification of some of the ten dimensions.  Thus the ten-dimensional spacetime
$X$ looks like the product $X= {\cal K}^d \times \IR^{1,9-d}$, with ${\cal K}$ a
suitable compact
manifold or orbifold.  Depending on which compact space is taken, it will be the
quantity of preserved supersymmetry. 

All five theories and their compactifications are parametrized by: the string coupling
constant $g_S$, the geometry of the compact manifold ${\cal K}$, the topology of ${\cal K}$ and the
spectrum of bosonic fields in the {\bf NS-NS} and the {\bf R-R} sectors.  Thus one can
define the {\it string moduli space} ${\cal M}$ of each one of the theories as the space
of all associated parameters. Moreover, it can be defined a map between two of these
moduli spaces. The dual map is defined as the map ${\cal S}: {\cal M} \to {\cal M}'$
between the moduli spaces ${\cal M}$ and ${\cal M}'$ such that the strong/weak
region of ${\cal M}$ is interchanged with the weak/strong region of ${\cal M}'$. 
One can define another map ${\cal T}:  {\cal M} \to {\cal M}'$ which
interchanges the volume $V$ of ${\cal K}$ for ${1 \over V}$. One example of the map ${\cal T}$
is the equivalence, by T-duality, between the theories Type IIA compactified on ${\bf
S}^1$ at radius $R$ and the Type IIB theory on ${\bf S}^1$ at raduis ${1 \over R}$.  The
theories HE and HO constitutes another example of the
${\cal T}$ map. In this section we will follows the Sen's review \senlec. Another
useful references are \refs{\em,\schwarz,\town,\vafa,\kirtwo}. Type IIB theory is self-dual
with
respect the ${\cal S}$ map. 

\vskip 1truecm

\noindent
{\it Type IIB-IIB Duality}

The Type IIB theory is self-dual. In order to see that write the bosonic part of the 
action of Type IIB supersting theory

$$
S_{\bf IIB} = {1 \over 2 \kappa^2} \int_X d^{10}x \sqrt{-G_{\bf IIB}} 
e^{-2\Phi} \bigg( R + 4 \partial_{I} \Phi \partial^{I} \Phi 
-{1 \over 2} H_{IJK} H^{IJK} \bigg)  
$$
\eqn\socho{
 -{1\over 4 \kappa^2} \int_X  d^{10}x \sqrt{-G_{IIB}}
\bigg( |F_{(1)}|^2 + |\widetilde{F}_{(3)}|^2
+ {1 \over 2}  |\widetilde{F}_{(5)}|^2 \bigg) - {1\over 4\kappa^2} \int_X A_{(4)} \wedge
H_{(3)} \wedge F_{(3)},}
where $\widetilde{F}_{(3)} = dA_{(2)} -  a \wedge H_{(3)}$ and  $\widetilde{F}_{(5)} =
dA_{(4)} -
{1 \over 2} A_{(2)} \wedge H_{(3)} + {1 \over 2} B_{(2)} \wedge F_{(3)}$. 

This action is clearly invariant under 
$$
\Phi' = - \Phi , \ \ \ \ \ \  G'_{IJ}= e^{-\Phi} G_{IJ},
$$
\eqn\snueve{
 B_{(2)} = A_{(2)} \ \ \ \ \ \  A'_{(2)}= - B_{(2)}, \ \ \ \ \ \  A'_{(4)} = A_{(4)}.} 
This symmetry leads to an identification
of a fundamental string F1 with a D1-brane ($B_{(2)} = A_{(2)}$) and the interchanging of
a pair of D3-branes.

\vskip 1truecm

\noindent
{\it Type I-SO(32)-Heterotic Duality}

In order to analyze the duality between Type I and SO(32) heterotic string theories we
recall the spectrum of both theories. These fields are the dynamical fields
of a supergravity Lagrangian in ten dimensions. Type I string theory has in the {\bf
NS-NS} sector the following fields: the metric $G_{IJ}^{\bf I}$, the dilaton $\Phi^{\bf I}$ and
in the {\bf R-R} sector: the antisymmetric tensor $A_{IJ}^{\bf I}$. Also there are 496
gauge bosons $A^{a{\bf I}}_{I}$ in the adjoint representation of the gauge group SO(32). For
the SO(32)  heterotic string theory the spectrum consist of: the spacetime metric
$G_{IJ}^{\bf H}$, the dilaton field $\Phi^{\bf H}$, the antisymmetric tensor $B_{IJ}^{\bf H}$
and 496 gauge fields $A^{a {\bf H}}_{I}$ in the adjoint representation of SO(32). Both
theories have ${\cal N} = 1$ spacetime supersymmetry. The effective action for the
massless fields of the Type I supergravity effective action $S_{\bf I}$ is defined as

$$
S_{\bf I} = {1 \over 2 \kappa^2} \int_X d^{10}x \sqrt{-G^{\bf I}} 
e^{-2\Phi^{\bf I}} \bigg( R + 4 (\nabla \Phi^{\bf I})^2  
-{1 \over 12} |\widetilde{F}_{(3)}|^2 \bigg)  
$$

\eqn\setenta{
- {1 \over g^2} \int_X d^{10}x \sqrt{-G^{\bf I}} e^{- \Phi^{\bf I}} Tr(F^{\bf
I}_{IJ} F^{{\bf I}IJ}) }
where $\widetilde{F}_{(3)} = F_{(3)} - {\alpha ' \over 4} [\omega_{3Y}(A) - \omega_{3L}
(\omega)].$  

While the heterotic action $S_{\bf H}$ is defined as 

\eqn\seuno{
S_{\bf H} = {1 \over 2 \kappa^2} \int_X d^{10}x \sqrt{-G^{\bf H}} 
e^{-2\Phi^{\bf H}} \bigg[ R + 4 (\nabla \Phi^{\bf H})^2 
-{1 \over 12} |\widetilde{H}_{(3)}|^2  - {\alpha ' \over 8} Tr ( F^{\bf H}_{IJ} F^{{\bf H}IJ}) 
\bigg]}
where $\widetilde{H}_{(3)} = dB_{(2)} -  {\alpha ' \over 4} [\omega_{3Y}(A) - \omega_{3L}
(\omega)].$

The comparison of these two actions leads to the following identification of the fields

$$
 G_{IJ}^{\bf I} = e^{-\Phi^{\bf H}} G_{IJ}^{\bf H}, \ \ \ \ \ \ \ \  \Phi^{\bf I} = - \Phi^{\bf
H},
$$

\eqn\sedos{
 A_{I}^{a{\bf I}} = A_{I}^{a{\bf H}}, \ \ \ \ \ \ \ \ \  \widetilde{F}^{\bf I}_{(3)} =
\widetilde{H}^{\bf H}_{(3)}.}
This give us many information, the first relation tell us that the metrics of both
theories are the same.  The second relation interchanges the $B_{(2)}$ field in the {\bf
NS-NS} sector and the $A_{(2)}$ field in the {\bf R-R} sector. That implies the interchanging of
heterotic strings
and Type I D1-branes.  The third relation identifies the gauge fields coming from the
Chan-Paton factors from the Type I side with the gauge fields coming from the 16
compactified internal dimensions of the heterotic string. Finally, the opposite sign for
the dilaton relation means that the string coupling constant $g^{\bf I}_S$ is inverted
$g^{\bf H}_S = 1/ g^{\bf I}_S$ within this identification, and interchanges the strong and weak
couplings of both theories leading to the explicit realization of the ${\cal S}$ map. 

\vskip 1truecm	
\subsec{M-Theory}

We have described how to construct dual pairs of string theories. By the uses of the
${\cal S}$ and the ${\cal T}$ maps a network of theories can be constructed in various
dimensions all of them related by dualities.  However new theories can emerge from this
picture, this is the case of M-theory. M-theory (the name come from `mystery', `magic',
`matrix', `membrane', etc.) was originally defined as the strong coupling limit for Type
IIA string theory \em. 

It is known that Type IIA theory can be
obtained from the dimensional reduction of the eleven dimensional supergravity theory (a
theory known from the 70's years) and given by the Cremmer-Julia-Scherk Lagrangian

\eqn\setres{
I_{11}  =  {1 \over 2 \kappa^2_{11}} \int_Y d^{11}x \sqrt{-G_{11}} \bigg(R - |dA_3|^2\bigg) 
- {1\over 6} \int_Y
A_{(3)} \wedge F_{(4)} \wedge F_{(4)},} 
where $Y$ is the eleven dimensional manifold. If we assume that the eleven-dimensional 
spacetime factorizes as  $Y= X \times {\bf S}^1_R$, where the compact dimension has 
radius $R$. Usual Kaluza-Klein dimensional reduction leads to get the ten-dimensional metric, 
an scalar field and a vector field. $A_{(3)}$ from the eleven  dimensional theory leads to
$A_{(3)}$ and $A_{(2)}$ in the ten-dimensional theory. The scalar field turn out too be proportional
to the dilaton field of the {\bf NS-NS} sector of the Type IIA theory. The vector field from the KK
compactification can be identified with the $A^{\bf IIA}$ field of the {\bf R-R} sector. From the
three-form in eleven dimensions are obtained the RR field $A_{(3)}$ of the Type IIA theory. Fin ally,
the $A_{(2)}$ field is identified with the NS-NS B-field of field strength $H_{(3)} = d B_{(2)}$. 
Thus the eleven-dimensional Lagrangian leads to the Type IIA supergravity in the weak coupling
limit ($\Phi \to 0$ or $R \to 0$). The ten-dimensional IIA supergravity  describing the bosonic
part of the low energy
limit of the Type IIA superstring theory is 

$$
S_{\bf IIA} = {1 \over 2 \kappa^2} \int_X d^{10}x \sqrt{-G^{\bf IIA}}
e^{-2\Phi^{\bf IIA}} \bigg( R + 4 (\nabla \Phi^{\bf IIA})^2 
-{1 \over 12} |H_{(3)}|^2 \bigg) 
$$

\eqn\secuatro{
- {1 \over 4 \kappa^2} \int_X d^{10}x \sqrt{-G^{\bf IIA}}
\bigg( |F_{(2)}|^2 +  
|\widetilde{F}_{(4)}|^2 \bigg)
- {1 \over 4 \kappa^2} \int_X  B_{(2)} \wedge dA_{(3)} \wedge  dA_{(3)}}
where ${H}_{(3)} = dB_{(2)}$, $F_{(2)} = dA_{(1)}$ and  
$\widetilde{F}_{(4)} = dA_{(3)} - A_{(1)} \wedge H_{(3)}.$	

It is conjectured that there exist an eleven dimensional
fundamental theory whose low energy limit is the 11 dimensional supergravity theory
and that it is the strong coupling limit of the Type IIA superstring theory.
At the present time the degrees of freedom (dof's) are still unknown, through at the
macroscopic level they should be membranes and fivebranes (also called M2-branes and
M5-branes).

\vskip 1truecm
\subsec{Horava-Witten Theory}

Just as the M-theory compactification on ${\bf S}^1_R$ leads to the Type IIA theory,
Horava and Witten realized that orbifold compactifications leads to the $E_8 \times E_8$
heterotic theory in ten dimensions HE (see for instance \town). More precisely

\eqn\secinco{
{\rm M}/({\bf S}^1/\IZ_2) \Longleftrightarrow HE}
where ${\bf S}^1/ \IZ_2$ is homeomorphic to the finite interval $I$ and the
$M$-theory is thus defined on $Y = X\times I$. From the ten-dimensional point of
view, this configuration is seen as two parallel planes placed at the two boundaries
$\partial I$ of $I$. Dimensional reduction and anomalies cancellation conditions imply
that the gauge degrees of freedom should be trapped on the ten-dimensional planes $X$
with the gauge group being $E_8$ in each plane.  While that the gravity is propagating
in the bulk and thus both copies of $X$'s are only connected gravitationally.

\vskip 1truecm
\subsec{F-Theory}

$F$-Theory was formulated by C. Vafa, looking for an analog theory to M-Theory for
describing non-perturbative compactifications of Type IIB theory (for a review see
\refs{\vafa,\senlec}). Usually in perturbative compactifications the parameter $\lambda =
a + i exp(-\Phi/2)$ is taken to be constant. $F$-theory generalizes this fact by
considering variable $\lambda$. Thus $F$-theory is defined as a twelve-dimensional
theory whose compactification on the elliptic fibration $T^2 - {\cal K} \to B$, gives
the Type IIB theory compactified on $B$ (for a suitable space $B$) with the
identification of $\lambda(\vec{z})$ with the modulus $\tau(\vec{z})$ of the torus
$T^2$. These compactifications can be related to the $M$-theory compactifications
through the known $S$ mapping ${\cal S}: IIA \to M/{\bf S}^1$ and the ${\cal T}$ map
between Type IIA and IIB theories. This gives

\eqn\seseis{
F/{\cal K}\times {\bf S}^1 \Longleftrightarrow M/{\cal K}.}
  Thus the spectrum of massless states of $F$-theory compactifications can be
described in terms of $M$-theory. Other interesting $F$-theory compactifications
are the Calabi-Yau compactifications

\eqn\sesiete{
 F/CY \Leftrightarrow  H/K3.}

\vskip 6truecm
\newsec{Non-perturbative Calabi-Yau Compactifications}

\noindent
{\it M-theory Vacua}

In this section we review some Calabi-Yau compactifications of M and F-Theories.  In the first
part of these lecture we described the perturbative CY compactifications, the purpose of the present
section
is see how these compactifications behaves in the light of duality and D-brane theory (for
excellent reviews see \refs{\andreas,\ovrut}). The presence of D-branes or M-branes, in the
case of M theory, modifies the perturbative CY compactifications, here we briefly
describe these modifications.

Assume that the eleven-dimensional spacetime is $Y = M \times {\bf S}^1/\IZ_2
\times {\cal K}$, with ${\cal K}$ being a Calabi-Yau three-fold. Here we consider that
${\cal K}$ is a elliptic fibration, since they are favored by CY compactifications of
M and F theories. These spacetime corresponds of having two copies (planes) of $X=M \times
{\cal K}$ at the two boundaries of the orbifold. According to the Horava-Witten theory, anomalies
cancellation involves that one ${\cal N}=1$ vector supermultiplet of the $E_8$ super
Yang-Mills theory has to be captured in each orbifold fixed plane $X_i$, $i=1,2$. 

According to the perturbative description it is necessary to specify now two stable or semi-stable
holomorphic vector
bundles $V_i$ on ${\cal K}$ with arbitrary group structure. For the heterotic-M theory
compactifications the structure group has to be a subgroup of $E_8$. 
For simplicity we restrict ourselves to SU$(n_i)$ vector bundles $V_i$ over ${\cal K}$. The presence
of fivebranes is of extreme importance here, since it allows more flexibility to construct such
vector bundles $V_i$ which leads to more realistic particle physics models. From the modified
Bianchi identity and the anomaly cancellation condition of the orbifold system and the
fivebranes wrapped on holomorphic two-cycles of ${\cal K}$, leads that these bundles are subject to
the cohomological constraint of the second Chern
classes $c_2(V_1) + c_2(V_2) + [W] = c_2(T{\cal K})$, where $[W]$ is the topological class
associated to the fivebranes. 

The description of the low-energy physics requires of the computation of the first three
Chern classes of the holomorphic bundles $V_i$ over ${\cal K}$ and thus determine completely 
a {\it non-perturbative vacuum}. M and F theories compactifications 
require that ${\cal K}$ must be a holomorphic elliptic fibration. Thus the construction of these
bundles are nontrivial.

\vskip 1truecm
\noindent
{\it Construction of the Gauge Bundles over Elliptic Fibrations}

An holomorphic elliptically fibered Calabi-Yau three-fold is a fibration

$$ 
{\cal K} \buildrel{\pi}\over{\to} B
$$
where $B$ is an auxiliary complex two-dimensional manifold, $\pi$ is an holomorphic mapping,
and for each $b \in B$, $\pi^{-1}(\{b\})$ is isomorphic to an elliptic curve $E_b$. In
addition we require the existence of a global section $\sigma: B \to {\cal K}$ of this
fibration. 

The elliptic fibration can be characterized by a single line bundle ${\cal L}$ 
over $B$, ${\cal L} \to B$, whose fiber is the cotangent space to the elliptic curve, $T^*E_b$. This
bundle satisfies the condition: ${\cal L} = K_B^{-1}$ with $K_B$ being the
canonical bundle of $B$, under the usual condition that the canonical bundle 
$K_{\cal K}$ has vanishing first Chern class $c_1(K_{\cal K})=0$. While the global section is
specified giving the bundles $K_B^{- \otimes 4}$ and  $K_B^{- \otimes 6}$. 

These conditions are known to be satisfied by
base spaces $B$ corresponding to {\it del Pezzo, Hirzebruch} and {\it Enriques} surfaces.

For elliptic fibrations, Friedman, Morgan and Witten \fmw\ found that the second Chern class
of 
the holomorphic tangent bundle $T{\cal K}$ can be written in terms of the Chern classes of $B$ as
follows

\eqn\seocho{
c_2(T{\cal K}) = c_2(B) + 11 c_1^2(B) + 12 \sigma c_1(B),}
where $c_1(B)$ and $c_2(B)$ are the first and the second class of $B$ and $\sigma$ is a two-form
and represents the
Poincar\'e dual of mentioned global section of the elliptic fibration.

One can construct the semi-stable SU$(n_i)$ holomorphic bundles $V_i$ on ${\cal K}$ through 
the specification of two line bundles
$\widehat{\cal L}$ with first Chern class $\eta \equiv c_1(\widehat{\cal L})$ and $\widehat{\cal W}$
with corresponding first
Chern class $c_1(\widehat{\cal W})$ depending on some parameters $n, \sigma, c_1(B), \eta$ and
$\lambda$. Thus the bundle $\widehat{\cal W}$ is completely specified by the elliptic fibration and
the line bundle $\widehat{\cal L}$. The condition that $c_1(\widehat{\cal W}) \in \IZ$
leads to the relation $\lambda = m +{1 \over 2}$ for $n$ odd and $\lambda = m $
and $\eta = c_1(B)$ mod 2, for $n$ even, $m \in \IZ$. Thus the Chern classes of the
SU$(n)$ gauge bundle $V$ are

\item{$\bullet$} $c_1(V) = 0$

\item{$\bullet$} $c_2(V) = \eta \sigma - {1 \over 24}c_1^2(B)(n^3 - n) + {1 \over 2}
\big(\lambda^2 - {1 \over 4} \big) n \eta \big(\eta -n c_1(B)\big)$

\item{$\bullet$} $c_3(V) = 2 \lambda \sigma \eta \big(\eta - n c_1(B)\big).$

In order to construct realistic particle physics models we take a given base space $B$
and compute its corresponding Chern classes $c_1(B)$ and  $c_2(B)$.  Compute the 
relevant Chern classes of the SU$(n)$ gauge bundles $V$. The constraints above reduce the 
number of consistent physical non-perturbative vacua. Given appropriate $\eta$ and
$\lambda$ one can determine completely the physical Chern classes.

\vskip 1truecm
\noindent
{\it Counting the Number of Families}

The number of families of leptons and quarks of the four-dimensional theory is related to the number
of zero modes of the Dirac operator
coupled to gauge field,  which is a connection
on the SU$(n)$ bundle $V$ on the 
elliptic fibered Calabi-Yau three-fold ${\cal K}$. In order to count the number of the
chiral fermionic zero modes, one can consider the following cases:

\item{$\bullet$} $SU(3) \times E_6 \subset E_8:$ 
${\bf 248} = ({\bf 8},{\bf 1}) \oplus ({\bf 1},{\bf 78}) \oplus ({\bf 3},{\bf 27})
\oplus({\bf 3}^*,{\bf 27}^*).$

\item{$\bullet$} $SU(4) \times SO(10) \subset E_8:$ 
${\bf 248} = ({\bf 15},{\bf 1}) \oplus ({\bf 1},{\bf 45}) \oplus ({\bf 4},{\bf 16})
\oplus({\bf 4}^*,{\bf 16}^*).$

\item{$\bullet$} $SU(5) \times SU(5) \subset E_8:$ 
${\bf 248} = ({\bf 24},{\bf 1}) \oplus ({\bf 1},{\bf 24}) \oplus({\bf 10},{\bf 5})
\oplus({\bf 10}^*,{\bf 5}^*) \oplus ({\bf 5},{\bf 10}^*) \oplus ({\bf 5}^*,{\bf 10}).$

The matter representations appear in the fundamental representation of the gauge group
SU$(n)$. The index of the Dirac operator gives

\eqn\senueve{
\delta = index(\not \! \! D_{\cal K}) = \int_{\cal K} ch(V) td({\cal K}) = {1\over 2} \int_{\cal K}
c_3(V)}
where $td({\cal K})$ is the Todd class of ${\cal K}$. From explitit formula 
for $c_3(V)$ we get  that the number of generations is given by $\delta =
\lambda \eta \big( \eta - n c_1(B)\big)$.

\vskip 2truecm

\centerline{\bf Acknowledgements}
We are very grateful to the organizers of the {\it Ninth Mexican
School on Particles and Fields} for the opportunity to give these lectures. One
of us O. L-B. is supported by a CONACyT graduate fellowship. It is a pleasure to
thank A. G\"uijosa, A. P\'erez-Lorenzana, F. Quevedo, N. Quiroz, M. Ruiz-Altaba, and A.M. Uranga 
for very useful discussions and enjoyable collaboration.



\listrefs

\end